\documentclass{aa}
\usepackage{graphicx}
\usepackage{natbib}
\usepackage{txfonts}
\begin{document}
   \title{Desorption rates and sticking coefficients for CO and N$_2$ interstellar ices}

   \subtitle{}

   \author{S.~E. Bisschop \inst{1} \and H.~J. Fraser\inst{2} \and
          K.~I. \"{O}berg\inst{1,3} \and E.~F. van Dishoeck\inst{1} \and
          S.Schlemmer\inst{4} }

   \offprints{S.Bisschop} \mail{bisschop@strw.leidenuniv.nl}
   \institute{Raymond and Beverly Sackler Laboratory for Astrophysics at Leiden Observatory, Postbus 9513, 2300 RA Leiden, Netherlands
         \and
{Department of Physics, University of Strathclyde, 107 Rottenrow East, Glasgow G4 ONG, Scotland}
             \and
{Division of Geological and Planetary Sciences, California Institute
of Technology, Mail Stop 150-21, Pasadena, CA 91125, USA}
\and 
{I. Physikalisches Institut, Universit\"{a}t zu K\"{o}ln, Zulpicher Strasse 77, 50937 K\"{o}ln, Germany}
}
   \date{Received <date> / Accepted <date>}

\abstract{We present Temperature Programmed Desorption (TPD)
experiments of CO and N$_2$ ices in pure, layered and mixed
morphologies at various ice ``thicknesses'' and abundance ratios as
well as simultaneously taken Reflection Absorption Infrared Spectra
(RAIRS) of CO. A kinetic model has been developed to constrain the
binding energies of CO and N$_2$ in both pure and mixed environments
and to derive the kinetics for desorption, mixing and segregation. For
mixed ices N$_2$ desorption occurs in a single step whereas for
layered ices it proceeds in two steps, one corresponding to N$_2$
desorption from a pure N$_2$ ice environment and one corresponding to
desorption from a mixed ice environment. The latter is dominant for
astrophysically relevant ice ``thicknesses''. The ratio of the binding
energies, $R_{\rm BE}$, for pure N$_2$ and CO is found to be 0.936
$\pm$ 0.03, and to be close to 1 for mixed ice fractions.  The model
is applied to astrophysically relevant conditions for cold pre-stellar
cores and for protostars which start to heat their surroundings.  The
importance of treating CO desorption with zeroth rather than first
order kinetics is shown.
The experiments also provide lower limits of 0.87 $\pm$ 0.05 for the
sticking probabilities of CO-CO, N$_2$-CO and N$_2$-N$_2$ ices at 14
K. The combined results from the desorption experiments, the kinetic
model, and the sticking probability data lead to the conclusion that
these solid-state processes of CO and N$_2$ are very similar under
astrophysically relevant conditions. This conclusion affects the
explanations for the observed anti-correlations of gaseous CO and
N$_2$H$^+$ in pre-stellar and protostellar cores.

}

\authorrunning{Bisschop et al.}
\titlerunning{Desorption rates and sticking coefficients for CO and N$_2$ ices}
\maketitle
%
%________________________________________________________________

%__________________________________________________________________

\section{Introduction}
\label{intro}

CO and N$_2$ are two of the most abundant species in molecular clouds and therefore control the abundances of many other molecules. CO is the second most abundant molecule after H$_2$, both in the gas
phase and in the solid state. Gaseous CO abundances up to 2.7$\times$
10$^{-4}$ with respect to H$_2$ are found in warm regions
\citep{lacy1994}, indicating that CO contains most of the carbon not
locked up in refractory material. In cold clouds, CO ice
absorption features are seen superposed on the spectra of background
sources or embedded protostars \citep[e.g.,][]{chiar1994,pontoppidan2003}.  The
solid CO abundance varies strongly from source to source, but can be
as high as $10^{-4}$ with respect to H$_2$ in the coldest cores
\citep{pontoppidan2005}. Such high abundances are consistent with
indirect determinations of the amount of CO frozen out in the densest
parts of pre-stellar cores based on submillimeter line and continuum
data, which suggest that more than 90\% of the CO is removed from the
gas \citep[e.g.,][]{caselli1999,tafalla2004,jorgensen2005}.

The amount of N$_2$ present in the gas and solid state is more
uncertain, since N$_2$ cannot be detected directly as it
lacks a permanent dipole moment. The abundance of gas phase N$_2$
is usually inferred from the presence of the daughter species
N$_2$H$^+$. Early work by \citet{womack1992} inferred gas
phase N$_2$ abundances of 2--6 $\times 10^{-6}$ with respect to
H$_2$ in star-forming regions, indicating that N$_2$ contains at
most 10\% of the nitrogen abundance.  Up
to an order of magnitude higher abundances were found \citet{dishoeck1992}, suggesting that at least
in some sources the transformation to molecular form is complete. More recent determinations of the N$_2$ abundance have
focused on dark cores for which the physical structure is well
determined from complementary data. For example, \citet{bergin1995}
and \citet{bergin2002} find typical gas-phase N$_2$ abundances of
$1-2\times 10^{-5}$.  Indirect indications for N$_2$ freeze-out onto grains can be obtained from analysis of the millimeter N$_2$H$^+$ data, which suggest a
decline in the gas-phase abundance by a at least a factor of two in
the centers of dense cores \citep{bergin2002,belloche2004}. Constraints on the amount of solid N$_2$ that might be present come from analysis of the solid CO band profile \citep{elsila1997}. The most stringent limits indicate that the
N$_2$:CO ratio must be less than 1:1, derived for sources for which
both $^{12}$CO and $^{13}$CO ices have been detected
\citep{boogert2002,pontoppidan2003}.  This limit only holds for
mixed ices of CO and N$_2$, not when N$_2$ ice has formed a separate
layer. 

The chemistries of CO, N$_2$ and their daughter products are
intimately linked, even though the two molecules belong to different
elemental families. This is due to the fact that CO is one of the
main destroyers of N$_2$H$^+$ in the gas phase. When CO is frozen
out onto the grains, N$_2$H$^+$ is enhanced, as confirmed
observationally by the anti-correlation of the abundances of
N$_2$H$^+$ with CO and HCO$^+$ in pre- and protostellar regions
\citep{bergin2001,tafalla2002,difranco2004,pagani2005,jorgensen2004}.
This anti-correlation is often quantitatively explained by a factor of 0.65 difference in the
binding energies for CO and N$_2$, allowing N$_2$ to
stay in the gas phase while CO is frozen out. These models do not contain an active grain-surface chemistry, but only include freeze-out and desorption. The relative
freeze-out behavior of CO and N$_2$ also affects the abundance of
H$_3^+$ and its level of deuterium fractionation
\citep{roberts2002}. Indeed, observations of H$_2$D$^+$ in cold
cores and in protoplanetary disks often invoke large (relative)
depletions of CO and N$_2$ \citep{ceccarelli2005}.

The above discussion clearly indicates the need for a good
understanding of the processes by which CO and N$_2$ freeze-out and
desorb from the grains under astrophysically relevant conditions. To describe desorption, accurate values for the binding
energies and the kinetics of the process are needed. For freeze-out,
the sticking probability is the main uncertainty entering the
equations.  In an earlier paper \citep[][hereafter paper I]{oberg2005}, we presented a limited set of experiments using our new
ultra-high vacuum (UHV) set-up to show that the ratio of the binding
energies $R_{\rm BE}$ for CO and N$_2$ in mixed and layered ices is
at least 0.923 $\pm$ 0.003 and in many circumstances close to unity.  This result can be understood chemically by the fact that the two molecules are
iso-electronic.  Indeed, the sublimation enthalpies calculated from the IUPAC
accredited data for pure ices were found to be 756 $\pm$ 5 K and 826
$\pm$ 5 K for pure N$_2$ and CO ices respectively, giving a ratio of
0.915 \citep{lide2002}.  This experimental ratio is much larger than
the value $R_{\rm BE}=0.65$ adopted in chemical models to explain
the observational data \citep{bergin1997,ceccarelli2005}. In an
alternative approach, \citet{flower2005} used the results from paper
I and instead varied the sticking probabilities of CO and N$_2$,
which were assumed to be 1 below 15 K in all previous
models. They could only reproduce the observed anti-correlation of
N$_2$H$^+$ and HCO$^+$ if the sticking probability for N$_2$ was
lowered to 0.1 compared with 1 for all other molecules. 

In this paper, we present new experiments on CO--N$_2$ ices, both in
pure, layered and mixed ice morphologies with varying ice ``thicknesses'' and relative abundances. In addition to TPD, RAIRS is used to probe the
mixing, segregation and desorption processes in the ices. The aim of
these experiments is to understand the CO--N$_2$ ice system to an
extent that the experimental desorption kinetics can be modeled and
reproduced, and to subsequently use these model parameters to
predict the behavior of CO and N$_2$ under astrophysically relevant
conditions. The key parameters to be derived for the CO--N$_2$ ice
are: i) the CO-CO, CO-N$_2$, and N$_2$-N$_2$ binding energies, ii)
the desorption kinetics (i.e., the desorption rates), iii) the
diffusion kinetics (i.e., the mixing and segregation rates), and iv) lower limits to the sticking probabilities.

This paper is organized as follows: Sect. \ref{expt} focuses on the
experimental procedure and choice of ice layers and mixtures, Sect.
\ref{results} presents the experimental results on desorption, Sect.
\ref{model} a kinetic model of the experimental data, and Sect.
\ref{sticking} experiments on the sticking probabilities.  Sect.
\ref{astro} discusses how the kinetic model can be applied to
astrophysically relevant situations and predicts the desorption behavior of CO and N$_2$ for astrophysically relevant heating rates. In Sect. \ref{conclusion}
all important conclusions are summarized.

\section{Experimental procedure}
\label{expt}

\begin{figure}
 \resizebox{\hsize}{!}{\includegraphics{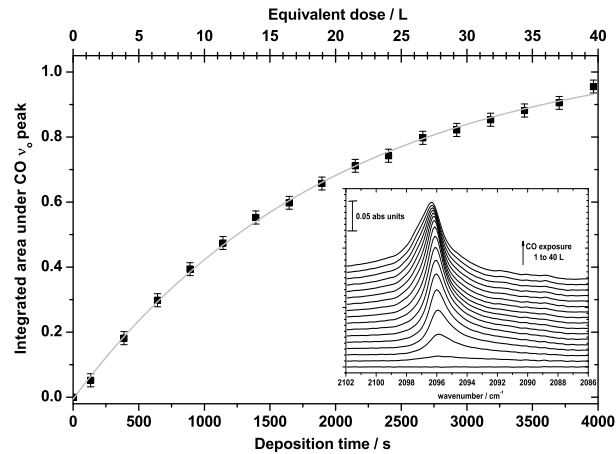}}
   \caption{Integrated intensity of the $^{13}$CO RAIR spectra with deposition time. Individual RAIR spectra are shown in the inset for ice exposures of 1 to 40 L in steps of 3 L (Langmuir).}
              \label{COgrowth}
    \end{figure}

The experimental apparatus used for this work, CRYOPAD (Cryogenic
Photoproduct Analysis Device) \citep{broekhuizen2005}, is very
similar to the SURFRESIDE Leiden surface astrochemistry instrument
described in detail elsewhere \citep{fraser2004b}. Briefly, all
experiments were performed in an ultra-high vacuum (UHV) chamber,
capable of reaching base pressures of better than 1 $\times$
10$^{-10}$ Torr.  At the center of the chamber is a gold-coated
copper substrate, mounted in close thermal contact with a closed
cycle He cryostat, which cools the whole substrate to 14 K. The
cryostat and substrate assembly is mounted on a rotation stage which
can be rotated through 360 deg. The sample temperature is controlled
to better than $\pm$ 0.1 K using the cryostat cold finger, a
resistive heating element and a Lakeshore 340 temperature control
unit. The system temperature is monitored with two KP-type (0.07\% Au
in Fe versus chromel) thermocouples, one mounted on the substrate
face, the second by the heater element. Ices are grown \emph{in
situ} onto the substrate, by exposing the cold surface to a steady
flow of gas, introduced into the chamber via an all metal flow
control valve, with a modified outlet directed at the substrate
center, along the surface normal. TPD is induced by heating the substrate (and ice
sample) at a steady rate of 0.0017 K s$^{-1}$, using a
linear heating ramp controlled by a positive feedback loop from the
Lakeshore instrument. The ice film is monitored using FT-RAIRS
(Fourier Transform RAIRS),
which is an analysis technique providing information on the
orientation and constituents of the ice film. The RAIR spectra
cannot be directly compared to observational data, however, since
they differ from transmission spectra. During flow setting,
deposition and desorption, gases liberated from the surface are
monitored using a quadrupole mass spectrometer (Pfeiffer Prisma).

To enable CO and N$_2$ to be discriminated from each other (and the
background signal) with mass spectrometry, isotopes of both
molecules were used, i.e. $^{13}$CO (Icon Isotopes 99.998\% m/e =
29), and $^{15}$N$_2$ (Cambridge Isotopes Inc. 98\% m/e = 30). This
isotopic substitution is simply an experimental asset and does not
affect the results presented in Sect.~\ref{results}: $^{12}$CO and
$^{14}$N$_2$ will behave identically. In the pure and layered ice
morphologies, the gases were used as supplied; to form the mixed
ices a 1:1 gas mixture of $^{13}$CO:$^{15}$N$_2$ was pre-prepared
and mounted on the UHV chamber gas-dosing system.  The dosing rate
for ice-film growth was set prior to cooling the sample, by
sequentially backfilling the chamber with the gas(es) of interest,
to a pressure of around 1 $\times$ 10$^{-8}$ Torr, equivalent to an
ion reading on the mass spectrometer of 7.5$\times10^{-10}$ A for
both $^{15}$N$_2$ and $^{13}$CO. The flow was then stopped, and the
background pressure within the chamber allowed to recover to
$\approx 1 \times 10^{-10}$ Torr, before the sample was cooled to 14
K. A background RAIRS spectrum was recorded prior to ice growth. The
ice films were then grown by reopening the pre-set flow valve for
exposure times equivalent to the gas dose required per sample gas
(see Table \ref{over_expt}), according to the morphology of the ice
to be grown, assuming 1 L (Langmuir) is $\approx$ 1 $\times$ 10$^{-6}$ Torr
s$^{-1}$, which roughly corresponds to $\sim$ 1 monolayer per unit area (cm$^{2}$) of material on the substrate. In the remainder of this paper, the ices are
discussed in terms of the gas exposure (in L) to which the substrate was subjected during ice-growth; for quick conversion to
astronomically relevant surface concentrations, it can be assumed
that a direct relationship exists between the ``exposure'' value
quoted, and surface coverage or ``thickness'' of the resulting ice,
which will be approximately $n$ monolayers of material, assuming an
exposure of $n$ L and a surface concentration of 10$^{15}$ molecules
cm$^{-2}$. 

During film growth, the CO-gas uptake on the cold surface
was monitored directly with RAIRS (see Fig. \ref{COgrowth}) and
indirectly by detecting residual CO and N$_2$ gas with the mass
spectrometer. Since N$_2$ has no permanent dipole, it is infrared
inactive and can only be monitored with the mass spectrometer. CO
ice growth was initially seen to be non-linear (Fig.
\ref{COgrowth}), most probably due to the preferential
formation of isolated 'islands' of CO on the substrate \citep[as is for example also seen by][]{nekrylova1993} rather than
an even, flat ``thin-film'' of CO-ice, where the substrate
surface is fully saturated. Around 40 L, CO ice growth becomes
linear, indicating that the structure of the ice that is forming no longer changes during deposition and the ice is present as a ``thin-film''. This is a key
reason for using an ice thickness of 40 L CO for experiments in
which the relative abundance of N$_2$ is varied. The ice was then heated in a TPD experiment (for a detailed discussion of
TPD experiments see e.g. \citet{menzel1982}), and 1 cm$^{-1}$ resolution
RAIR spectra were recorded as the
temperature reached $\approx$ 15, 20, 22, 24, 25, 26, 27, 28, 29,
30, 35 and 40 K.

The ice samples studied are summarized in Table \ref{over_expt}. Throughout this
paper, the notation X/Y indicates a layered ice morphology with X on
top of Y, whereas X:Y denotes a fully mixed ice system. The 1/1 and 1:1 notation denotes identical amounts of both species, whereas the x/40 L notation refers to experiments in which the ``thickness'' of the overlying N$_2$ layers is varied, but that of the CO layer is kept constant at 40 L.
The ``thicknesses'' have been chosen to be of astrophysical
relevance: if all condensible carbon were frozen out as
CO it would form an ice layer equivalent to $\sim$40
 monolayers on an interstellar grain \citep{pontoppidan2003}. This is a fortuitous coincidence with the point at which, experimentally, thin-film CO-ice growth
dominates in our apparatus. A layered ice morphology is indicated by
analysis of the interstellar solid CO profiles, which reveal a
component of pure CO ice which contains 60--90\% of the total solid
CO abundance and which is clearly separated from the H$_2$O ice
\citep{tielens1991,chiar1994,pontoppidan2003}.  Chemical models show that nitrogen is transformed into
N$_2$ at later times and at higher extinctions when
compared with the conversion of carbon from atomic form into CO
\citep{hendecourt1985,hasegawa1992}. Thus, either CO starts freezing
out before N$_2$ is formed so that N$_2$ forms a
``pure'' overlayer, or both molecules are present in the gas phase
and freeze out together. This makes N$_2$/CO and N$_2$:CO the most astrophysically relevant ice morphologies to study;
CO/N$_2$ ices were however also included in this study, to complete
our understanding of the behavior of the ice systems. In terms of relative abundances,
observational evidence (Sect. \ref{intro}) suggests that the N$_2$ abundance is always
less than or equal to that of CO. Models including gas-grain
chemistry predict N$_2$ ice abundances that are typically a factor
5--20 lower than those of CO ice
\citep{hasegawa1993,shalabiea1994,bergin1995,aikawa2005}. Together,
these arguments led to the choice of ice morphologies and exposures summarized in Table \ref{over_expt}.

\begin{table}
\caption{Overview of ice morphologies and ice exposure used in the experiments} \centering
\begin{tabular}{l|cc|c}
\hline
\hline
 & $^{13}$CO & $^{15}$N$_2$ & Total\\
 & L$^a$ & L$^a$ & L$^a$\\
\hline
Pure $^{13}$CO & 20 & - & 20\\
& 40 & - & 40$^b$ \\
& 80 & - & 80\\
\hline
Pure $^{15}$N$_2$ & - & 20 & 20\\
& - & 40 & 40$^b$\\
& - & 80 & 80\\
\hline
$^{13}$CO-$^{15}$N$_2$ & 10& 10 & 20\\
& 20 & 20 & 40\\
& 40 & 40 & 80$^b$\\
& 80 & 80 & 160\\
\hline
$^{13}$CO/$^{15}$N$_2$ & 10 & 10 & 20\\
& 40 & 40 & 80$^b$\\
& 80 & 80 & 160\\
\hline
$^{15}$N$_2$/$^{13}$CO & 10 & 10 & 20\\
& 20 & 20 & 40\\
& 40 & 40 & 80$^b$\\
& 80 & 80 & 160\\
& 5 & 40 & 45\\
& 10 & 40 & 50\\
& 20 & 40 & 60\\
& 30 & 40 & 70\\
& 50 & 40 & 90\\
\hline
\multicolumn{4}{l}{$^a$ in Langmuir (see Sect. \ref{expt})}\\
\multicolumn{4}{l}{$^b$ data previously reported in paper I}\\
\end{tabular}
\label{over_expt}
\end{table}

\section{Experimental results}
\label{results}

\begin{figure*}
\resizebox{\hsize}{!}{\includegraphics[width=8cm]{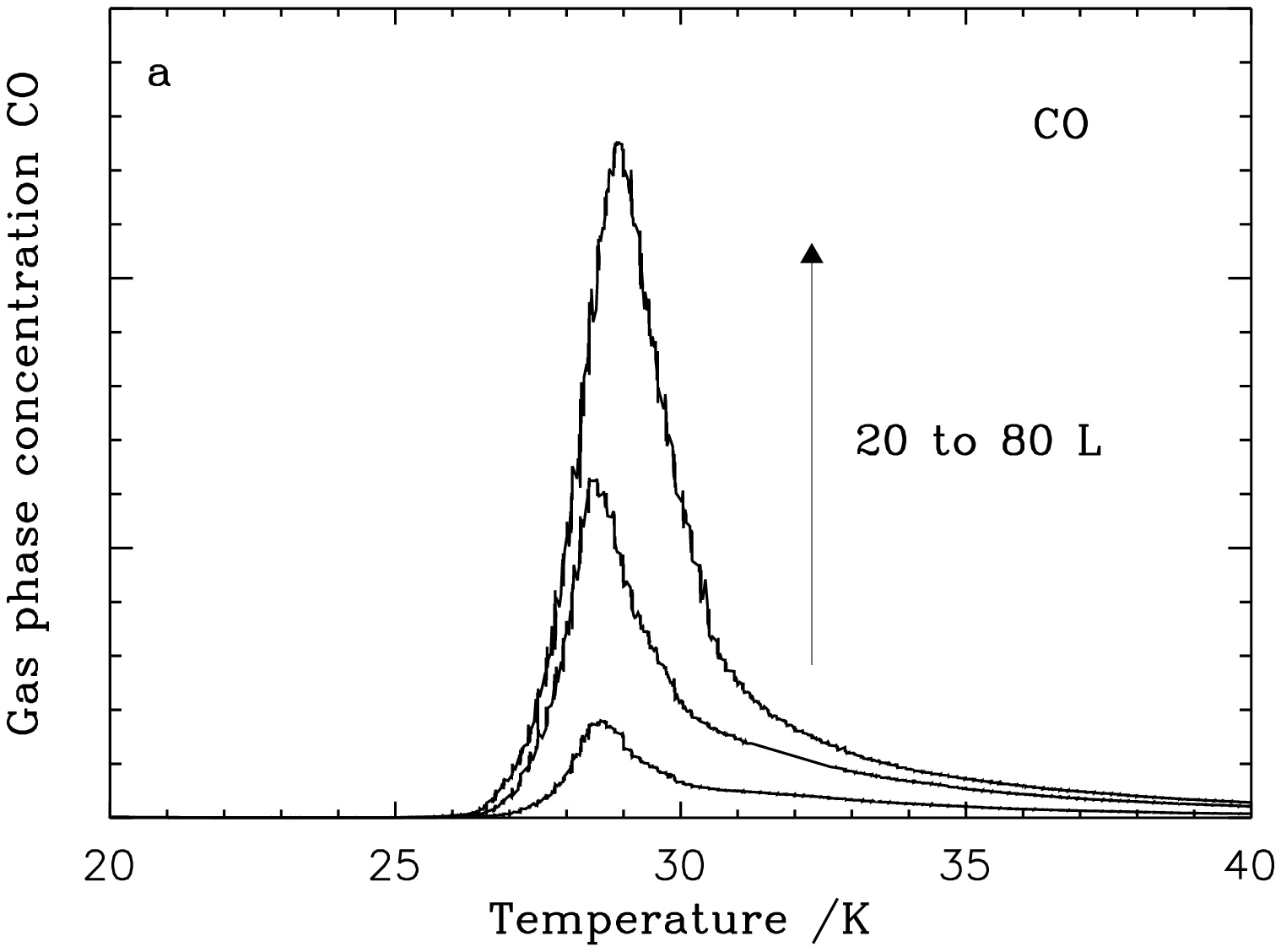}\includegraphics[width=8cm]{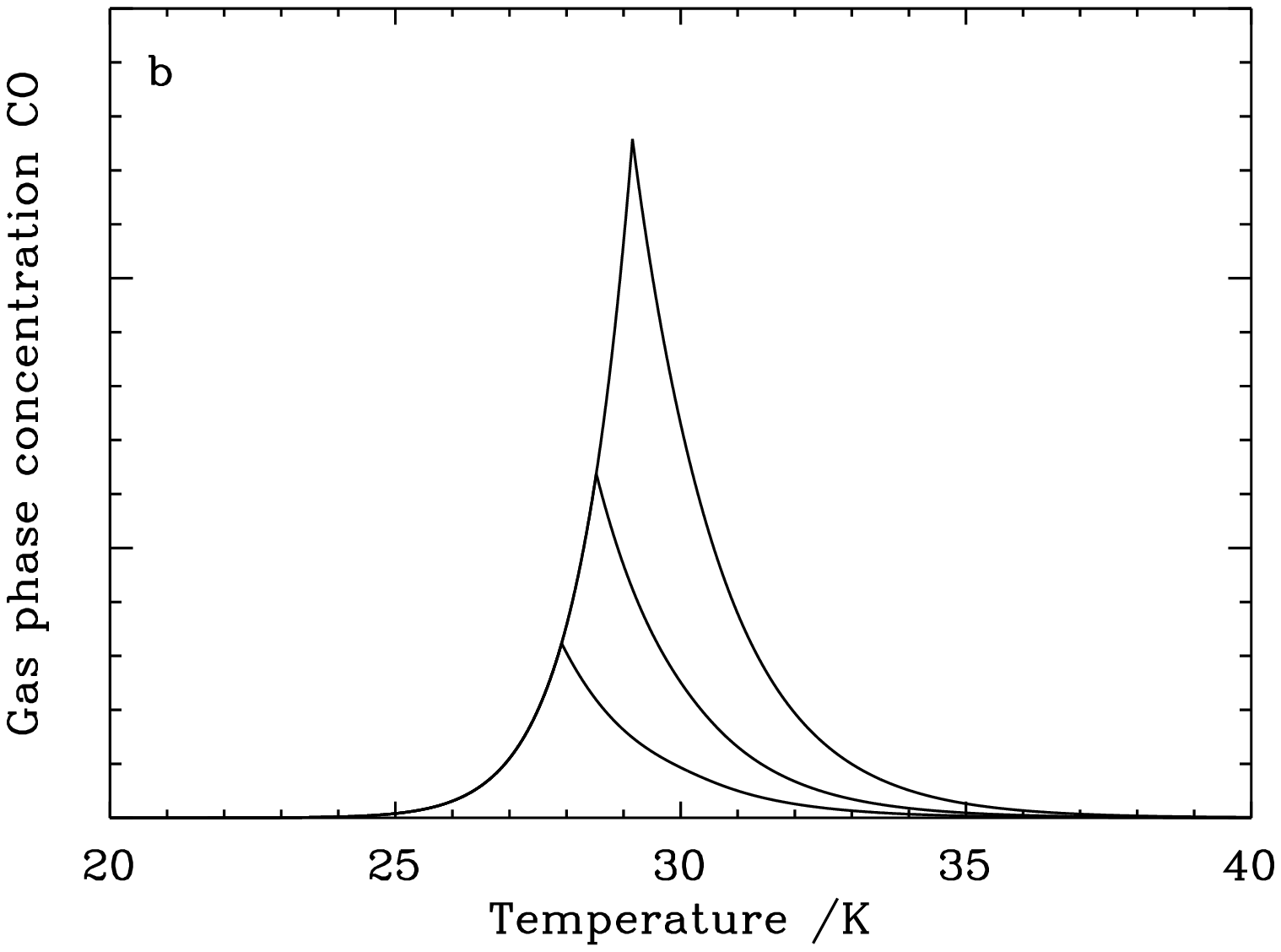}}
\resizebox{\hsize}{!}{\includegraphics[width=8cm]{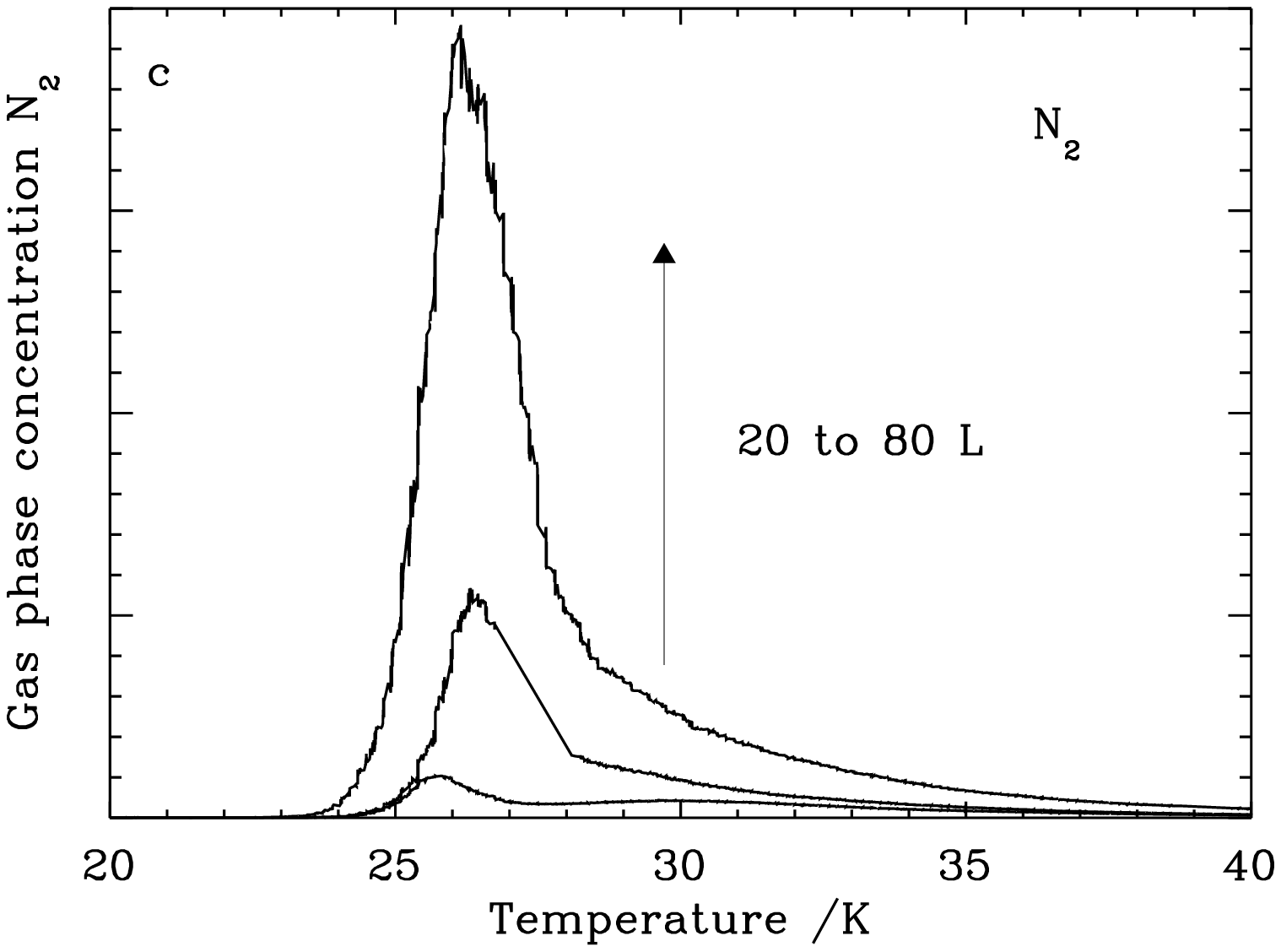}\includegraphics[width=8cm]{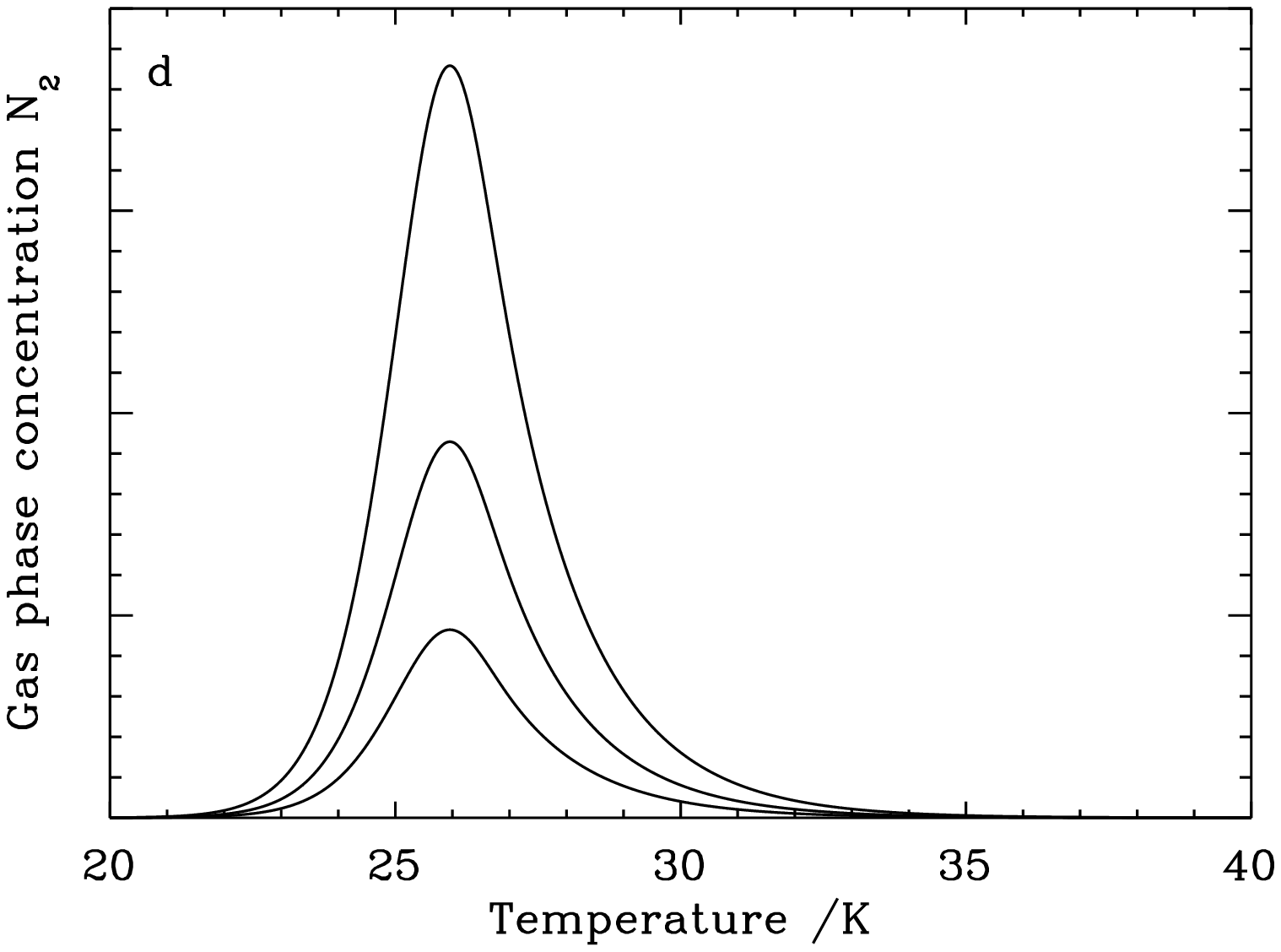}}
   \caption{TPD spectra for pure ices with exposures of 20, 40, 80 L. (a) CO experiments, (b) CO model, (c) N$_2$ experiments, and (d) N$_2$ model}
              \label{tpd_pure}
    \end{figure*}

\begin{figure*}
   \centering
   \includegraphics[width=23cm,angle=90]{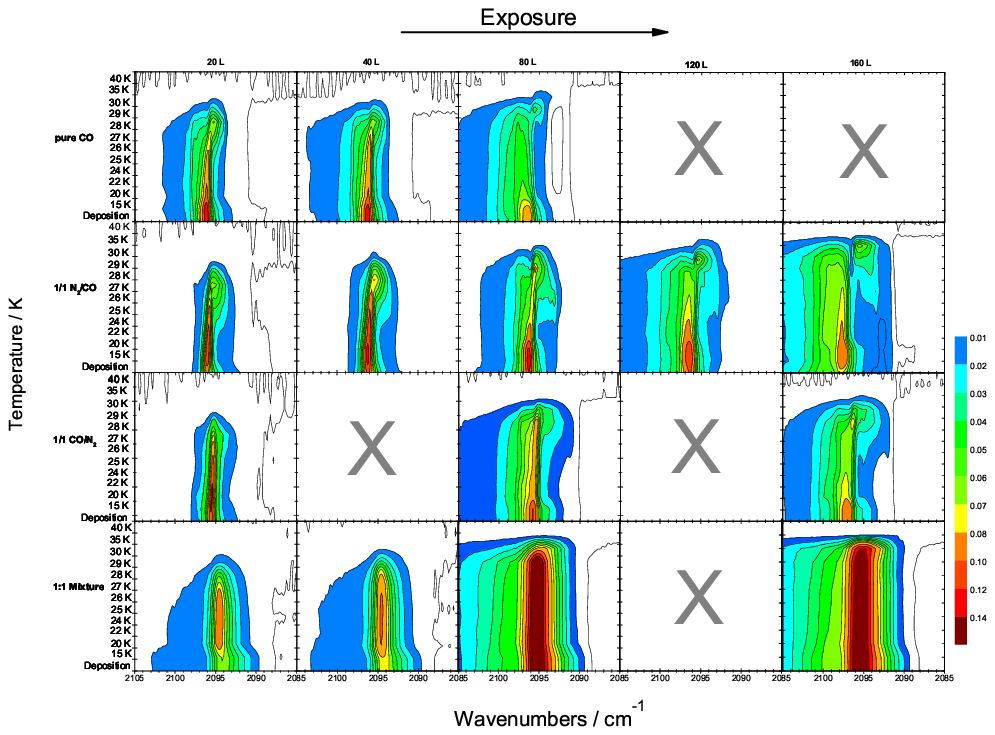}
   \caption{2-D RAIR spectra of $^{13}$CO plotted as frequency vs. temperature (on a non-linear temperature scale), in pure CO, mixed and layered CO:N$_2$ ices, with exposures indicated at the top of the matrix and ice morphologies on the left hand side. X= data not available.}
              \label{matrix_RAIRS}
    \end{figure*}

\subsection{Pure CO and N$_2$ ices}
\label{pure} In Fig. \ref{tpd_pure}a and c, the TPD spectra for
three different ice exposures, i.e. 20, 40, and 80 L, for pure CO
and N$_2$ ices are shown.  
The CO TPD curves indicate that the onset
for desorption is at around 26 K in the laboratory. The
leading edges of the TPD curves for the 40 and 80 L exposures
overlap, suggesting that the desorption process occurs
at a rate that is independent of the ice thickness. Consequently the
peak of the CO TPD curve shifts to higher temperatures for
increasing ice thicknesses, peaking at 28 K for an exposure of
40 L. This indicates the presence of multilayer
films, since the number of molecules that desorb depends only upon the number of molecules in the surface, which is identical at ice exposures of 40 and
80 L. Thus the desorption rate is constant until there are
no molecules left on the surface and desorption stops. This type of
kinetics is called zeroth order kinetics. The order of the kinetics is defined as the power of the number of molecules in the surface with which the rate of desorption scales (for details see Sect. \ref{construct_model}).  
Since the differences in the CO TPD spectra are smaller for all
ice morphologies, this is the only time they are discussed (Fig.
\ref{tpd_pure} a). The TPD signal for the 20 L experiment has a lower intensity than expected from scaling the 40 L data. This is due to island growth at low exposures (see Figure \ref{COgrowth} and Sect. \ref{expt}).

The onset of N$_2$ desorption shifts from 25 K for 20 and 40 L exposures,
to 24 K for 80 L (see Fig. \ref{tpd_pure}c). The peak position of
N$_2$ remains the same for the 40 and 80 L experiments. This
indicates that in contrast to CO, the desorption rate of N$_2$
increases with increasing ice thickness. This kind of kinetics is
called first order kinetics. Note that, in general, desorption
kinetics do not have to have an exact integer value. For
example \citet{bolina2005} find that multilayer desorption of
CH$_3$OH on highly oriented pyrolytic graphite (HOPG) has a
desorption order of 0.35. In most cases, however, the desorption
kinetics will approach either zeroth, first or even second order.

RAIRS data for pure $^{13}$CO 20, 40 and 80 L exposures are
shown in the first row of Fig. \ref{matrix_RAIRS}. The peak
position is around 2096 cm$^{-1}$ with a full width half maximum of
2 cm$^{-1}$. When the temperature increases above $\sim$20 K, a reduction in intensity and narrowing is observed on the blue
side of the CO band. This change is probably due to restructuring
of the ice. It is likely that the initial ballistic
deposition results in an ``open'' amorphous ice structure; at around
20 K the CO molecules become torsionally mobile about their lattice
points, resulting in an ``on the spot'' rotation about each molecule's
center of mass, and the formation of a more closely packed
structure. 
Finally, around 26 K
when pure CO ice starts desorbing, more dramatic changes occur in
the CO band. The origin of these changes is thought to be due to crystallization and is described in more detail in a future publication. The intensity decreases due to desorption, and
a small peak grows on the blue side of the main feature.

\subsection{Layered ices}
\label{layers}

\begin{figure*}
   \centering
\resizebox{\hsize}{!}{\includegraphics[width=8cm]{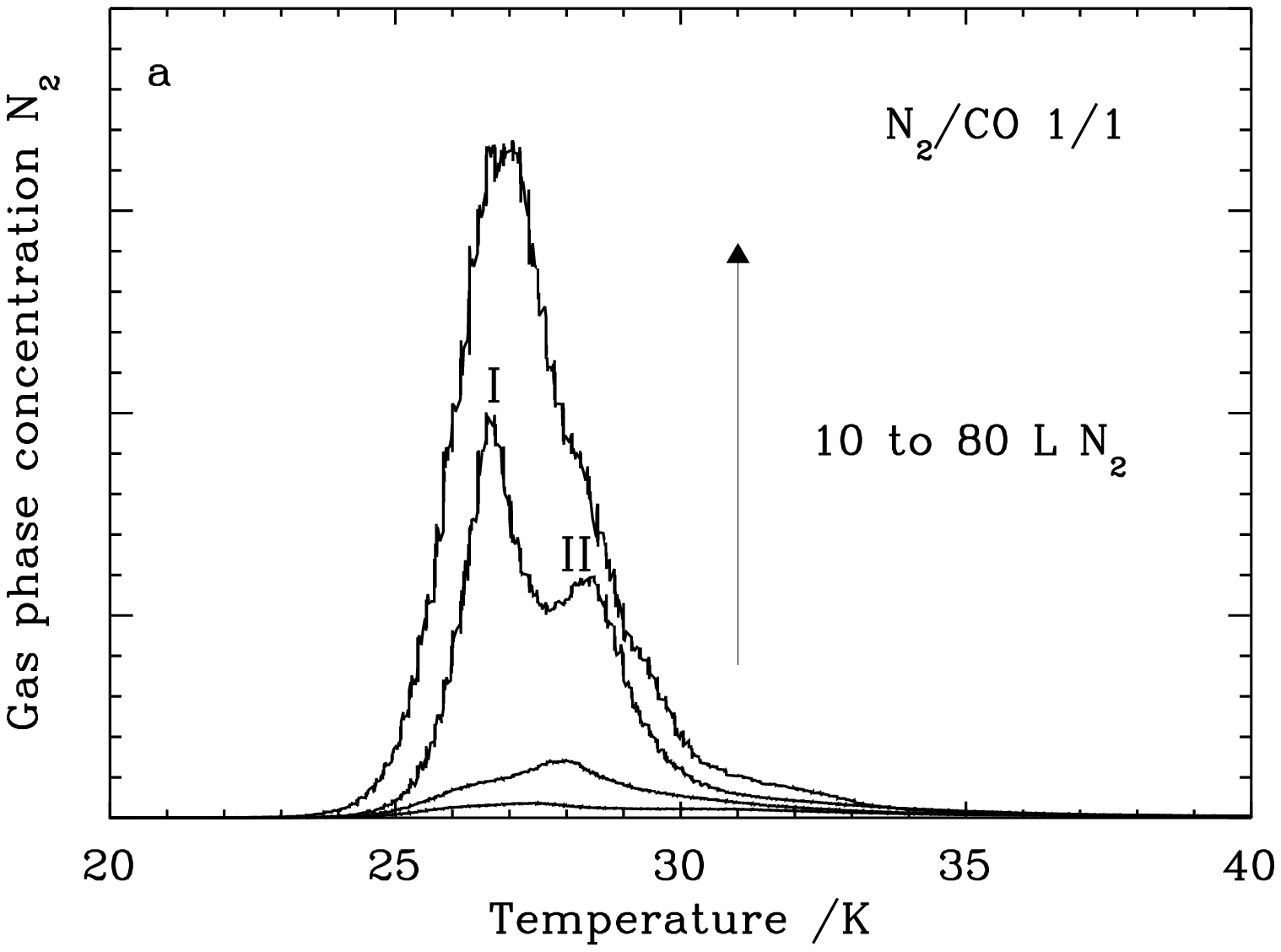}\includegraphics[width=8cm]{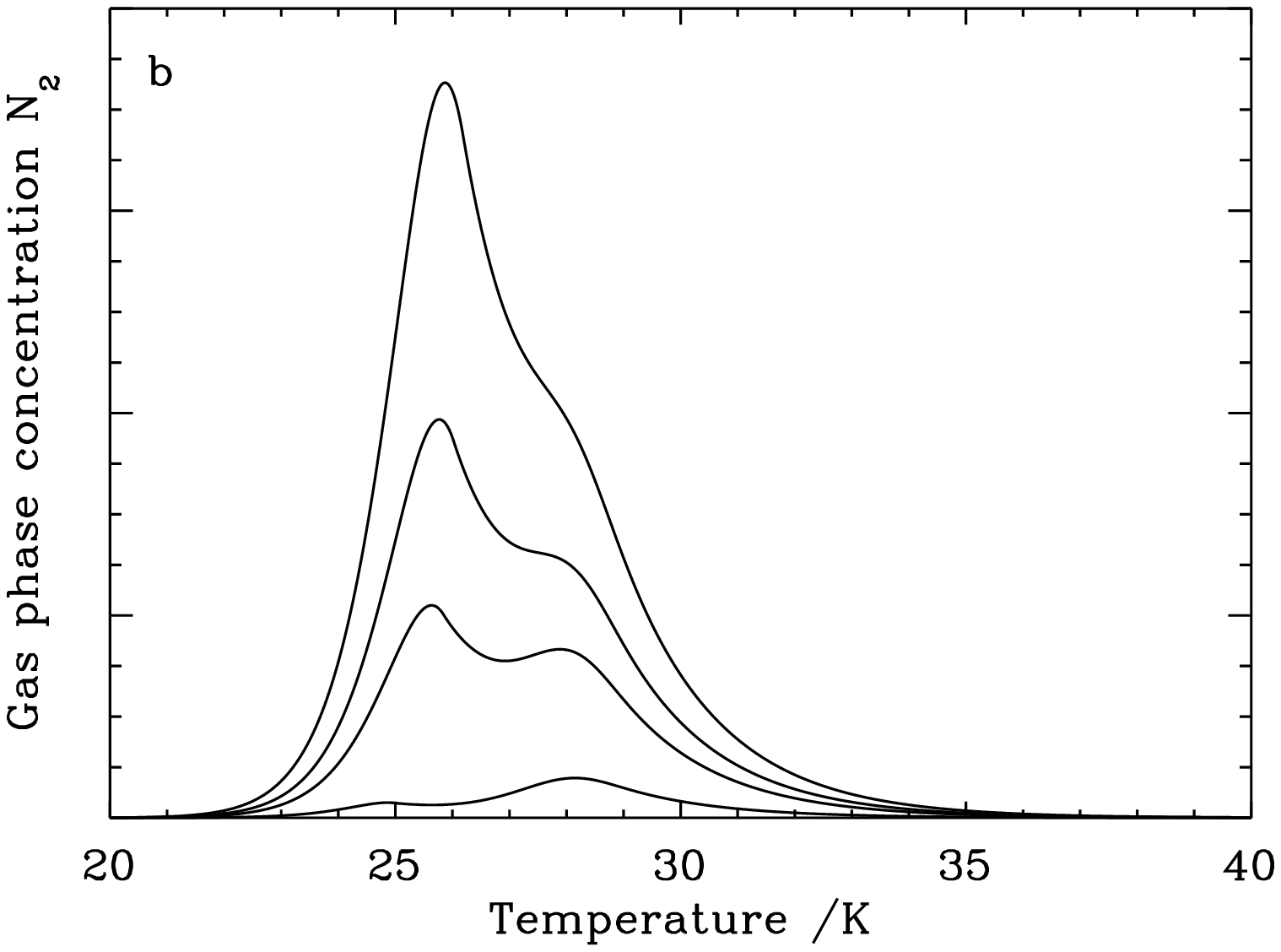}}
\resizebox{\hsize}{!}{\includegraphics[width=8cm]{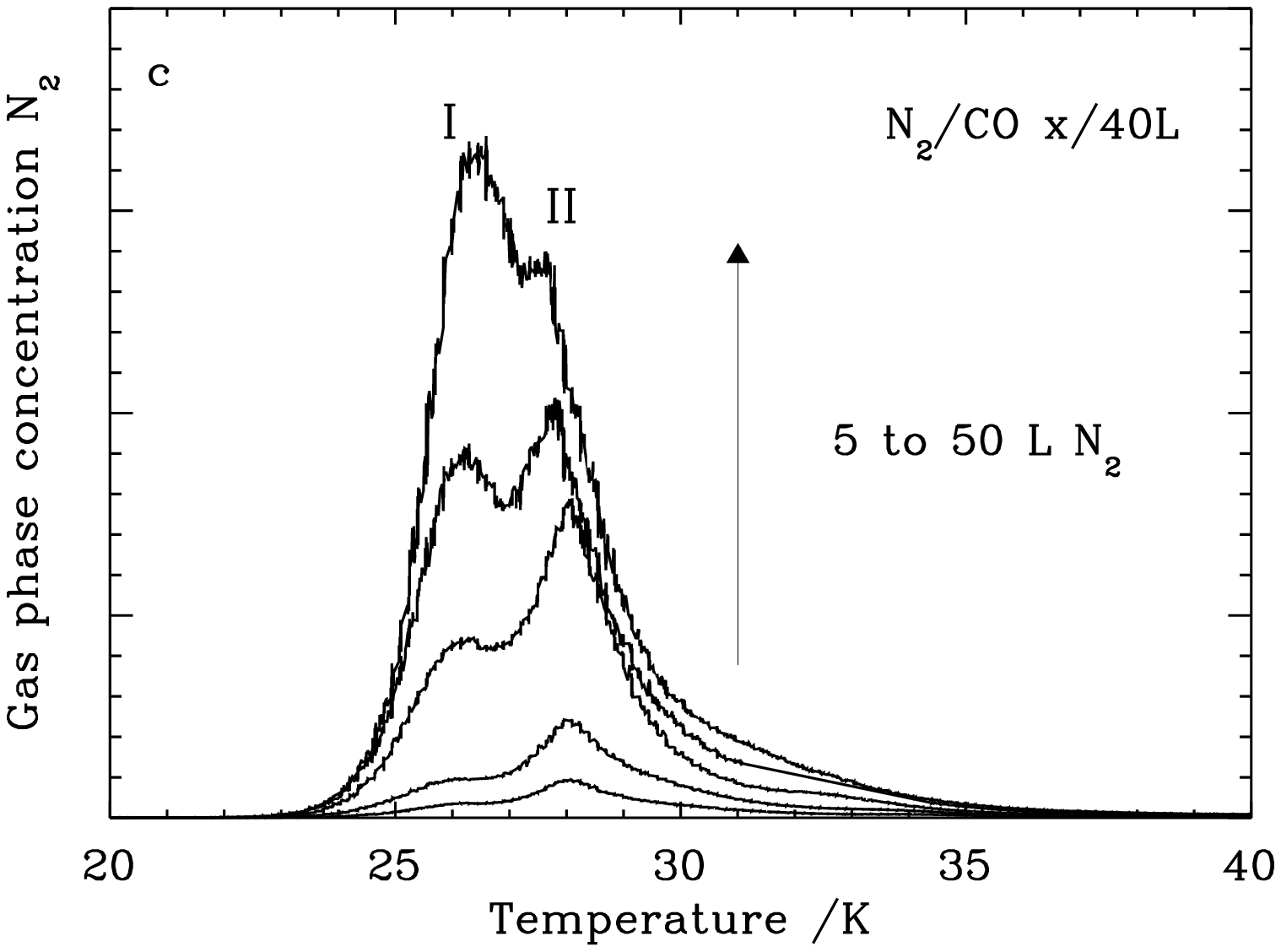}\includegraphics[width=8cm]{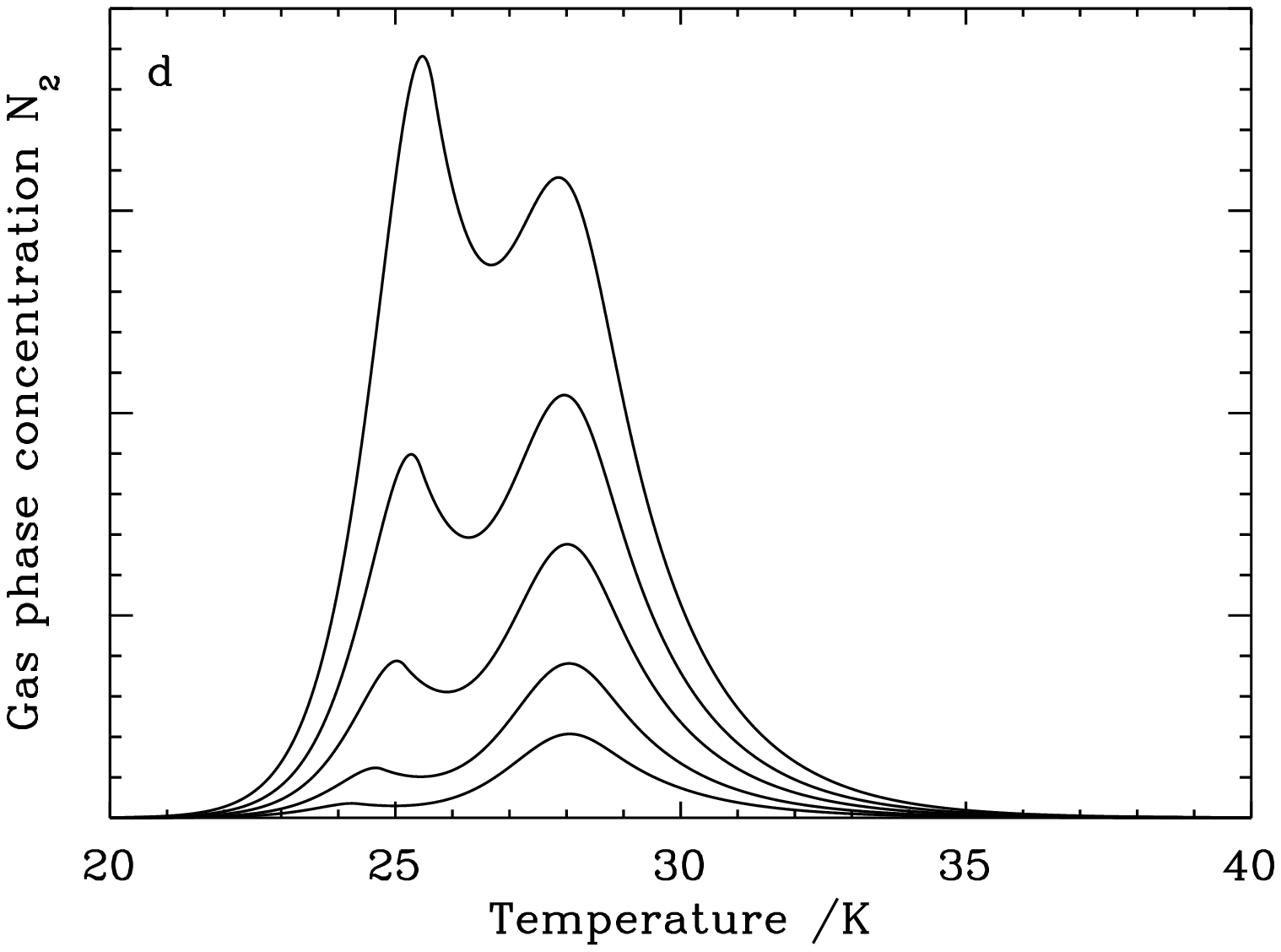}}
\resizebox{\hsize}{!}{\includegraphics[width=8cm]{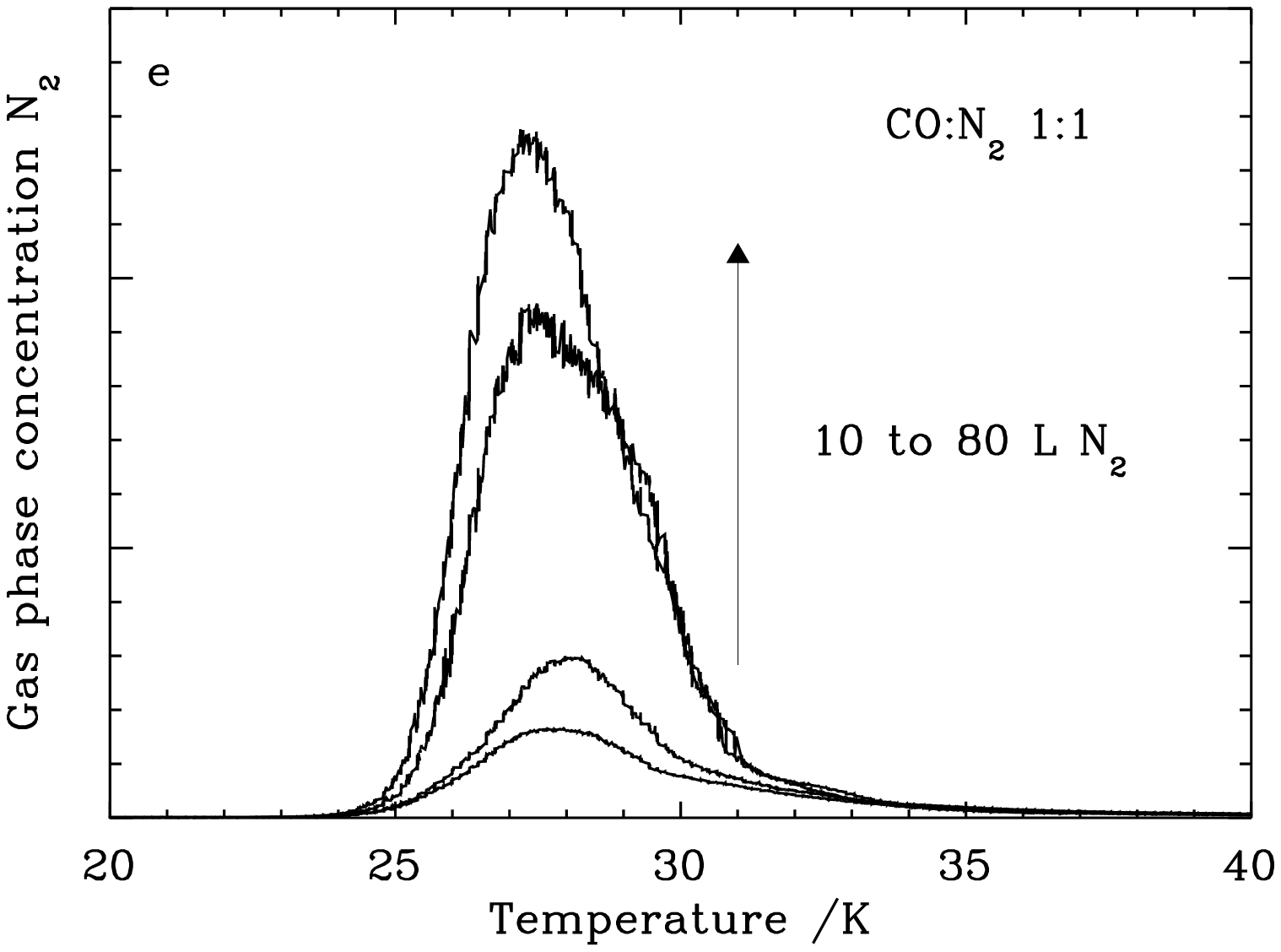}\includegraphics[width=8cm]{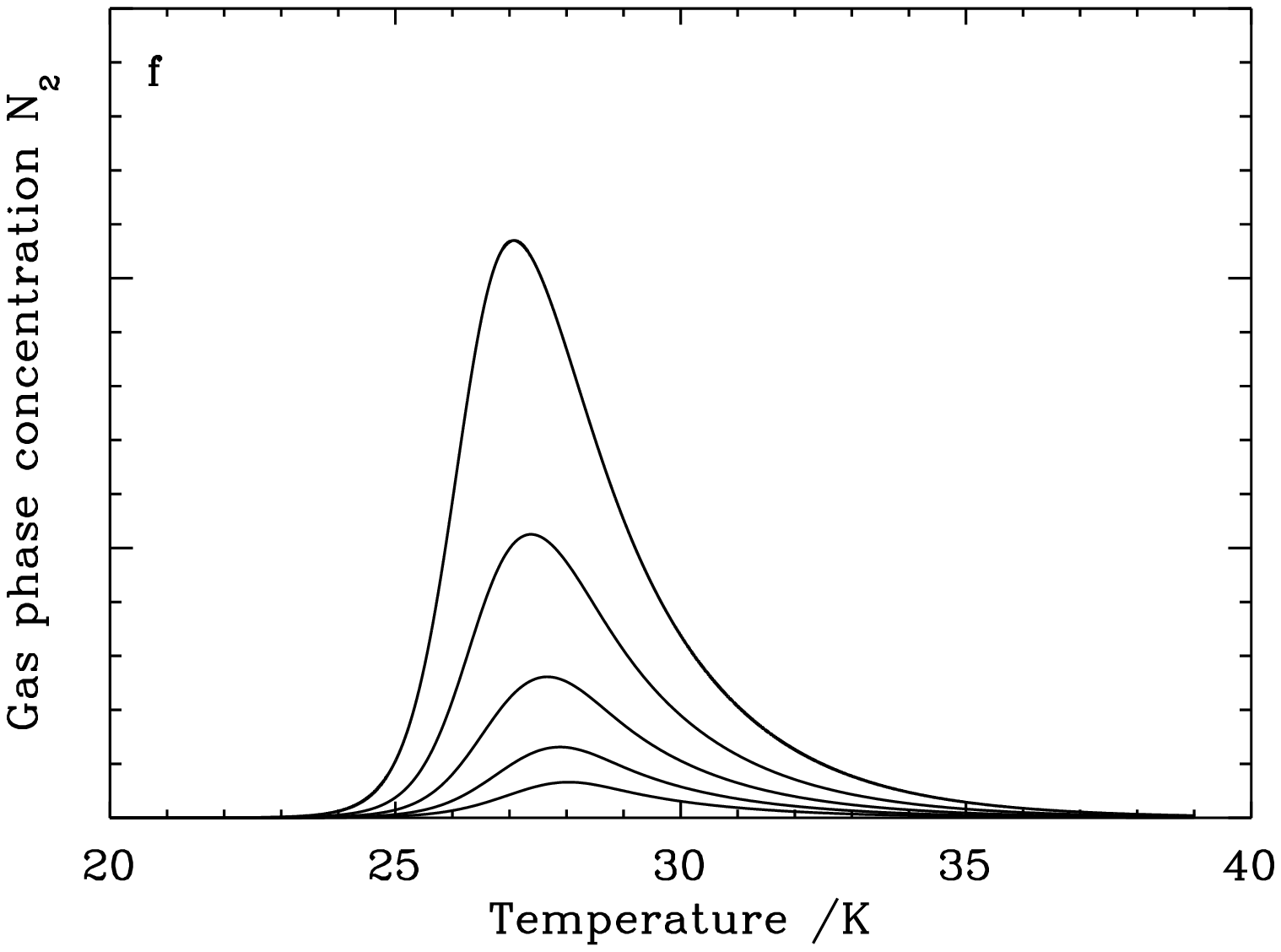}}
   \caption{N$_2$ TPD spectra: (a) (10-20-40-80 L)N$_2$/(10-20-40-80 L)CO, 1/1
layer, (c) (5-10-20-30-50 L)N$_2$/(40 L)CO, differential layer, e) (10-20-40-80
L)N$_2$:(10-20-40-80 L)CO, mixed ice 1:1. The equivalent model spectra are
shown in b, d, and f, respectively. The two experimental TPD peaks
are labeled I and II, corresponding to desorption of N$_2$ from
pure and mixed ice phases respectively.}
\label{tpd_overview}
\end{figure*}

\begin{figure}
   \centering
\resizebox{\hsize}{!}{\includegraphics{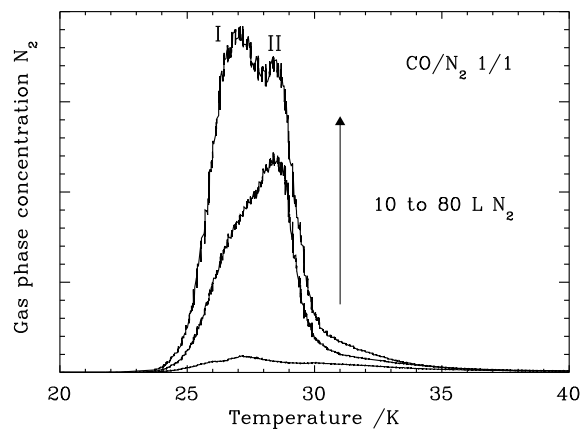} }
   \caption{N$_2$ TPD spectra of (10-40-80 L)CO/(10-40-80 L)N$_2$, 1/1 layer. The two experimental TPD peaks are labeled I and II, corresponding to desorption of N$_2$ from pure and mixed ice phases respectively.}
              \label{tpd_coonn2}
    \end{figure}

The N$_2$ TPD spectra for the 1/1 N$_2$/CO experiments  and
x/40 L N$_2$/CO are shown in Fig. \ref{tpd_overview}a and c
respectively. Additionally, the 1/1 experiments of CO/N$_2$ are
shown in Fig. \ref{tpd_coonn2}. In all cases at least one
peak is observed in the TPD spectra, but from the majority of the data
it is evident that the TPD spectra are actually composed from two
peaks, one at around 26 K (labeled peak I) and one at around 28 K
(labeled peak II). In all the spectra, peak I coincides with the
position of the TPD desorption peak in pure N$_2$, so it is
attributed to N$_2$ desorbing from a pure N$_2$ layer; peak II
coincides with the position of the pure CO TPD desorption peak and
is therefore assigned to co-desorption of N$_2$ with CO,
hypothesizing this occurs from a mixed phase of CO-N$_2$ ice. The
formation of this mixture would require bulk diffusion of N$_2$ and
 / or CO between the two separate layers. This mobility is found to commence at
significantly higher temperatures than those expected for the
hopping process on surfaces \citep{tielens1987}. The energy-barrier to hopping is typically assumed to be 0.3 $\times$ the binding energy, corresponding to around $\sim$ 285 K for CO and N$_2$ and implying that CO and N$_2$ are mobile around 10 K. Our much higher temperature for mobility is probably due to a much larger barrier to bulk diffusion than for surface diffusion . For comparison, experiments by \citet{collings2003b} suggest CO molecules become mobile at around 12-15 K on both CO and H$_2$O-ice surfaces, suggesting the
barrier to surface diffusion is only slightly higher than the theoretical
approximation used in astrochemical models.
Furthermore it is clear that the mixing process occurs during ice annealing, and not immediately on deposition, first because
there is significant N$_2$ desorption from a pure ice phase and
second because the desorption profiles of the layered and mixed ice
systems differ significantly (see Sect. \ref{mixed}).

Important information about the CO-N$_2$ ice system can be
derived from the relative intensities of peak I and II. In the 1/1
and x/40 L N$_2$/CO experiments, a turnover is observed between the peak intensities  (see Fig. \ref{tpd_overview}a and c), with peak II being more intense
than peak I for low ``thickness'', and visa versa at high ``thickness''. This turn-over occurs between the 40/40 L and 60/60 L
exposures for the 1/1 experiments, and between the 30/40 L and
50/40 L in the x/40 L experiments, i.e. both sets of experiments
consistently have the turn-over point around 40/40 L.

The CO RAIR spectra of the layered ices (second and third
rows of Fig. \ref{matrix_RAIRS}) have a $^{13}$CO feature that is
almost identical to that for pure $^{13}$CO, although the red-wing is less
pronounced. As for pure CO, the intensity of the blue-wing
decreases around 20 K, where the ice restructures, and a new peak
grows around 26 K, where CO starts to desorb. Since the changes in the layered ice spectra at 20 K are commensurate with
similar changes in the pure CO ice spectra, this is unlikely to be an indicator of the mixing process. Additionally, a blue
wing appears around 24-25 K, concurrently to the onset of N$_2$
desorption in the TPD spectra (see Fig. \ref{tpd_overview}). This feature is probably due to mixing of both molecules, as will be discussed in Sect. \ref{results_model}. The appearance of a blue wing around 24 K rather than 20 K reaffirms that the mixing process relies on bulk rather than surface diffusion.

Finally, the TPD spectra of 1/1 CO/N$_2$ ice layers at
exposures of 20, 80 and 160 L are shown in Fig. \ref{tpd_coonn2}.
These experiments were used primarily to test whether the ices were
indeed grown as separate layers on top of each other. The turn-over
point where peak I becomes more intense than peak II occurs at
slightly higher exposures compared with N$_2$/CO ice
layers, i.e. between 40/40 L and 80/80 L. It is therefore clear that
N$_2$ desorption is retarded by the CO overlayer, desorbing only
after it has mixed with, and (a fraction of which has) subsequently
segregated from, the CO-ice. As the spectra do not resemble those of
the pure N$_2$ ice, these experiments provide positive evidence that
the layer growth is sequential and coincident on the substrate. However, this ice structure is not thought to be astrophysically relevant, so is not discussed
further in this article.

\subsection{Mixed ices} 
\label{mixed}

The N$_2$ TPD spectra for mixed ices (Fig \ref{tpd_overview}e) differ from those of pure or layered ices in
that only one peak is observed, skewed to the low, and not high
temperature side of the desorption range. As the ``thickness'' of the
mixed ice increases, the TPD peak maximum shifts from 28 to 26 K.
This behavior indicates that at low exposures, N$_2$
desorbs predominantly from a mixed-ice environment, whereas as the exposure increases, a more significant fraction of
the N$_2$ is able to desorb form a pure N$_2$ layer. Furthermore,
the TPD peaks are broadened with respect to those observed
for pure and layered ice morphologies (see Sect. \ref{pure} and
\ref{layers}). This broadening is likely to be due to the merging of
peaks I and II, and the potential for a wider range of binding
environments to exist in the intimately mixed ice morphology. Desorption occurring from a pure N$_2$ ice environment suggests that
segregation must also occur within mixed CO-N$_2$ ice systems,
including the mixed phases that are formed in the layered ice
systems. However the fact that some desorption from the mixed phase is
always observed indicates that the segregation happens at a lower
rate than the mixing process, potentially because the energy barrier
to segregation is greater than that for mixing. This would
suggest that over certain temperature ranges the mixed ice phase is
thermodynamically more stable than the segregated layers.

The RAIR spectra of the mixed ices (final row Fig.
\ref{matrix_RAIRS}) differ from those of the pure and layered
ices, being broader (4 cm$^{-1}$) and shifted to 2094 cm$^{-1}$,
reflecting, as with the TPD data, that the structure of the mixed
ices is unique. Again, the CO band changes shape at around 20 K,
possibly due to a similar restructuring as observed for the pure and
layered ices, discussed in Sect. \ref{pure} and \ref{layers}, but
no further changes are observed as the temperature increases until
the ice starts desorbing. This implies that all or most of the CO remains in a mixed ice phase until it starts to desorb; even if
the concentration of this phase changes slightly as the N$_2$
segregates and desorbs, it is not evident in the RAIR spectra.

\section{Empirical model of CO-N$_2$ desorption} 
\label{model}

A model was built to gain a clearer qualitative
and quantitative understanding of the thermal annealing processes
including diffusion, mixing and desorption of the ices. The aims of this model are twofold; to reproduce the
experimental data and then apply the same kinetic parameters to
astrophysically relevant ice morphologies, temperatures and heating
rates.

\subsection{Constructing the model}
\label{construct_model}

The kinetic processes for desorption, mixing and segregation
in this system have a reaction barrier (i.e. they are
thermodynamically limited) and can therefore be described by
the following equation:

\begin{equation} r_{\rm des} = \frac{dN}{dt} =
\nu_i [N_s]^i e^{-E/T} \label{time_rate} \end{equation}

\noindent where $r_{\rm des}$ is the desorption rate
(molecules cm$^{-2}$ s$^{-1}$), $N$ is the number of molecules evaporating from the substrate 
(assuming throughout the remainder of these calculations that the
substrate has unit surface area (cm$^{2}$)), $t$ is time in s, $\nu_i$ the
pre-exponential factor (molecules$^{1-i}$ cm$^{2(i-1)}$ s$^{-1}$), $i$ is the reaction order, $[N_s]$ is the number of molecules partaking in a particular reaction per unit surface area (molecules cm$^{-2}$), $E$ is the
reaction barrier in K, which for the desorption processes can be
read as the binding energy, and $T$ is temperature in K. The physical meaning of the pre-exponential factor $\nu_i$ depends upon the reaction order $i$. For a first order reaction it refers to the lattice vibrational frequency which is typically in the range 10$^{11}$ - 10$^{13}$ s$^{-1}$; for zeroth order desorption it consists of the product of the lattice vibrational frequency with the surface density of order 10$^{15}$ cm$^{-2}$. Depending on the type of reaction, the
reaction order $i$ can vary, taking positive, negative and any
non-integer real value. Both $E$ and $\nu_i$ depend in principle upon ``thickness''. However, this dependence is not thought to be large since no major changes are observed between the FTIR spectra at different coverages indicating that the intermolecular environments are very similar.

In order to calculate the temperature-dependent rate measured in the
TPD experiments, the following conversion needs to be made:

\begin{equation} \frac{dN}{dt} = \frac{dN}{dT} \frac{dT}{dt}
\label{temp_rate}
\end{equation}

\noindent where ${dN}$/${dT}$ is the temperature-dependent
rate (molecules cm$^{-2}$ K$^{-1}$), and ${dT}$/${dt}$ the TPD
heating rate (K s$^{-1}$). At each time step, a fraction of the
molecules that have evaporated into the gas phase will be removed by
the pump; subtracting this rate from the desorption into the
gas phase will reproduce the experimental conditions. The
pump-rate is given by:

\begin{equation} r_{\rm pump} = \frac{dN}{dt} = - \nu_{\rm pump} N(g)
\label{pump_rate} \end{equation}

\noindent in which $\nu_{\rm pump}$ is the pump constant in s$^{-1}$ and
$N(g)$ the number of molecules entering the gas phase having desorbed from a unit surface area. To ensure the equations balance, N(g) is given in molecules cm$^{-2}$, implying that the molecules actually occupy a unit volume. Combining equations (\ref{time_rate}), (\ref{temp_rate}), and
(\ref{pump_rate}), the experimental results can be simulated in a
simple way. The reactions are summarized in Table~\ref{input}.

\subsection{Constraining the model}
\label{model_constraints}

First, the reactions $h$ and $i$ given in Table \ref{input} plus the
pump constants $\nu_{\rm pump}$ for CO and N$_2$ were
constrained, by fitting a first-order exponential to the
pump-down curves of both CO and N$_2$ at 14 K, accounting for the
pumping effects of the turbo-pump and the cryostat in the
experiment. Note that the $\nu_{\rm pump}$ values shown in Table
\ref{input} are experimentally determined and consequently fixed for
further iterations of the kinetic model.

\begin{table*}
\caption{Rate equations for desorption, mixing, and segregation of
CO and N$_2$ in the CO-N$_2$ ice systems.} 
\centering
\begin{tabular}{l|ccccc}
\hline
\hline
 & Reaction & Rate equation & $\nu$ & $E$ & $i$\\
& & & (molecules$^{(1-i)}$ cm$^{2(i-1)}$ s$^{-1}$) & (K) & \\
\hline
a & CO(s) $\rightarrow$ CO(g) &  $\nu_0$e$^{-E/T}$ & 7.0 $\times$ 10$^{26 \pm 1,  a}$ & 855 $\pm$ 25 & 0\\
b & N$_2$(s) $\rightarrow$ N$_2$(g) & $\nu_1$[N(s)]$^i$e$^{-E/T}$ & 1.0 $\times$ 10$^{11 \pm 1}$ & 800 $\pm$ 25 & 1\\
c & CO(mix) $\rightarrow$ CO(g) &  $\nu_1$[CO(mix)]$^i$e$^{-E/T}$ & 7.0 $\times$ 10$^{11 \pm 1}$ & 930 $\pm$ 25 & 1\\
d & N$_2$(mix) $\rightarrow$ N$_2$(g) & $\nu_1$[N$_2$(mix)]$^i$e$^{-E/T}$ & 1.0$\times$ 10$^{12 \pm 1}$ & 930 $\pm$ 25 & 1\\
e & CO(s)$\rightarrow$ CO(mix) & $\nu_0$e$^{-E/T}$ & 5.0$\times$ 10$^{26 \pm 1}$ & 775 $\pm$ 25 & 0\\
f & N$_2$(s)$\rightarrow$ N$_2$(mix) & $\nu_0$e$^{-E/T}$ & 5.0$\times$ 10$^{26 \pm 1}$ & 775 $\pm$ 25 & 0\\
g & CO(mix) + N$_2$(mix) $\rightarrow$ CO(s) + N$_2$(s) & $\nu_2$[CO(mix)][N$_2$(mix)]e$^{-E/T}$ & 1.0$\times$ 10$^{-4 \pm 1}$ & 930 $\pm$ 25 & 2\\
h & CO(g) $\rightarrow$ CO(pump) & $\nu_{pump}$[CO(g)] & 1.0 $\times$ 10 $^{-3,  a}$ & - & 1 \\
i & N$_2$(g) $\rightarrow$ N$_2$(pump) & $\nu_{pump}$[N$_2$(pump)] & 8.2$\times$10$^{-4,  a}$ & - & 1 \\
\hline
\multicolumn{6}{l}{$^a$ Parameters are fixed according to experimental constraints see Sect. \ref{model_constraints}.}
\end{tabular}
\label{input}
\end{table*}

Next, the parameters associated with reactions $a$ and $b$,
desorption from pure ice environments, were constrained. Since the
binding energies for pure CO ice desorption found by
\citet{collings2003b} and paper I are identical within experimental
error, the CO binding energy was initially set to the same value
reported in paper I; $\nu$ was fixed at the value reported by
\citet{collings2003b}. For N$_2$, the desorption kinetics appear to
be first order (see Sect. \ref{pure}) and therefore the
pre-exponential factor $\nu$ was initially estimated to be somewhere
between 10$^{11}$-10$^{13}$ s$^{-1}$, then varied in order to obtain
the best fit to the experimental data. The N$_2$ binding energy was
initially set to the value reported in Paper I, but also allowed to
vary in iterations of the model. The final values of these parameters
are given in Table \ref{input} and the corresponding TPD models are
presented next to the experimental data in Fig. \ref{tpd_pure}b and
d.

Desorption from the mixed ice fraction was assumed to be first order. This is thought to be a good assumption since the rate of desorption depends on the number of molecules on the surface and this will change after each molecule desorbs. Initially, the binding energy for desorption from the mixed ice layer was taken to be the same as that of pure CO desorption, because peak II appears to occur at the same temperature as the desorption of pure CO. However, when running the model this value had to be increased to reproduce the experimental effect.

From the TPD spectra described in Sect. \ref{results} no direct
measurement of the mixing rates was possible. Mixing can, however,
be inferred from the presence of peak II in the layered ice
experiments. Assuming a simple, single step process, reactions $e$
and $f$ in Table \ref{input} describe the mixing, assuming both
molecules contribute equally to the process. Good physical
arguments can be made for modeling this process as zeroth, first or
second order kinetics, and all three processes were investigated
(see  Appendix \ref{Ap_A} for a more detailed discussion). The outcome is that
experimental data are best reproduced if the mixing process is
zeroth order. Since mixing occurs only at the interface between the
CO and N$_2$ ice layers this description makes physical sense. A CO
or N$_2$ molecule at the interface has a certain chance of
overcoming the ``mixing barrier'' and diffusing into the opposite
layer, but the molecules remaining at the interface will still see
the same number of molecules, regardless of whether
there are 20 or 80 L of ice above or below it.

The final reaction to constrain is the segregation reaction
$g$. The relative number of molecules desorbing from pure N$_2$
environments in mixed ice morphologies increases with exposure, as
was discussed in Sect. \ref{mixed}. Segregation is modeled as one
reaction, in a second order process depending on the initial number of molecules in the mixed ice phase for both species. In reproducing all
the mixed ice experiments these values were, of course, equal. In the layered ices, however, it is unlikely that the relative abundances of
CO and N$_2$ in the mixed ice phases are equal. Consequently,
equation $g$ suggests that segregation is fastest from a equimolar
ice, decreasing as the relative abundances of either species deviate
from 1:1. Finally, from the TPD spectra of the layered ices it was
clear that the mixing was more efficient than the segregation
process, so, as discussed in Sect. \ref{layers}, the $E$ of reaction $g$ was always assumed to be greater than $E$ of reactions $e$ and $f$.

\subsection{Results}
\label{results_model}

In Table \ref{input}, all the model equations and best fit parameters ``by eye'' are given after running a large number of models. The error-bars arise from (i) the range of values over which simultaneous fits of $\nu$ and $E$ gave degenerate solutions to the model, (ii) the uncertainty in the number of molecules present on the surface, and (iii) the experimental uncertainties in the temperature. It is important to realize that the degeneracy in the simultaneous fits of $\nu$ and $E$ means that the combination of these values is more accurate than the individual values. Thus, in astrochemical models both parameters need to be used in combination to accurately reproduce the behavior of CO and N$_2$.

\begin{figure}
   \centering
\resizebox{\hsize}{!}{\includegraphics{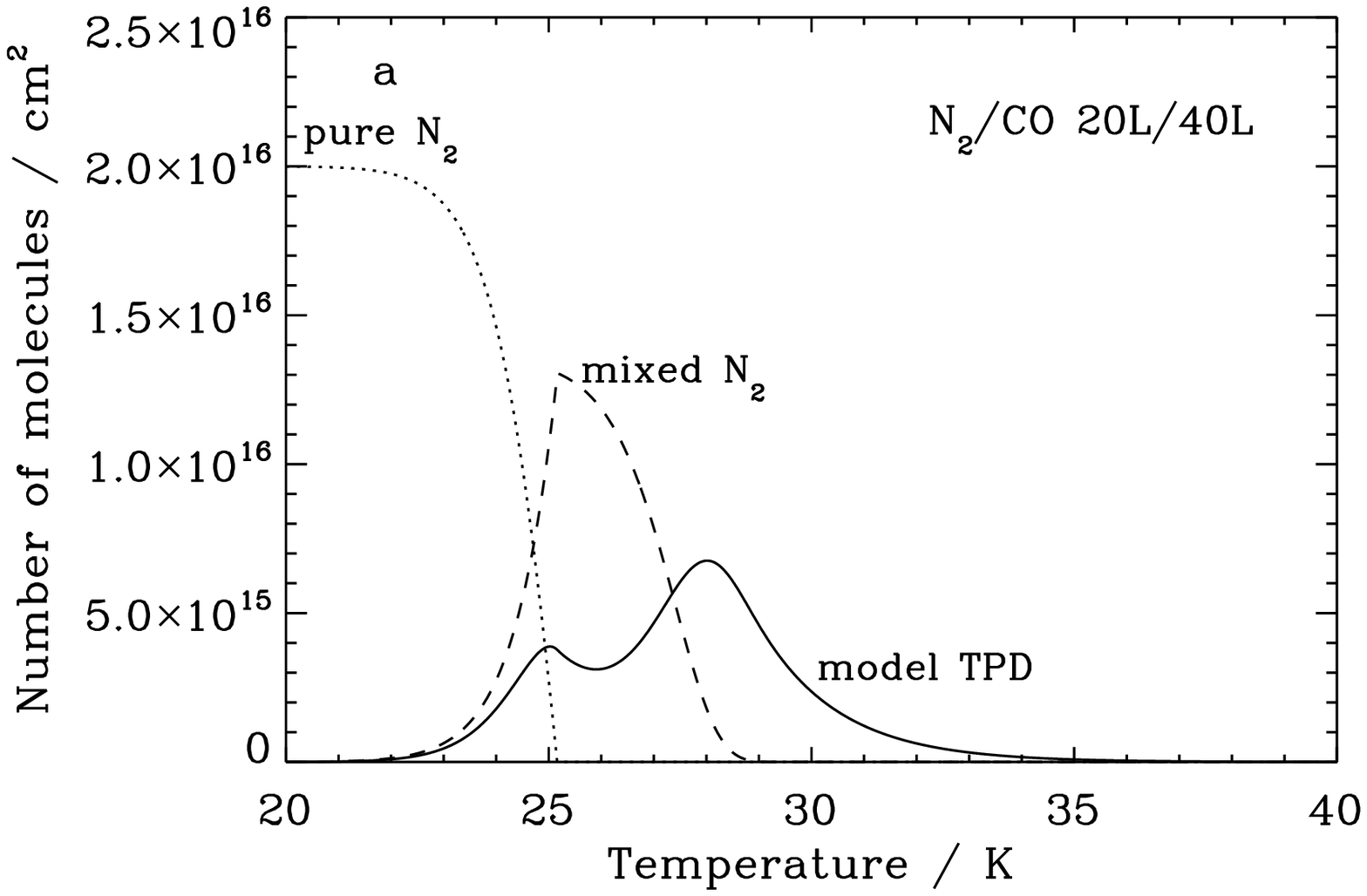}}
\resizebox{\hsize}{!}{\includegraphics{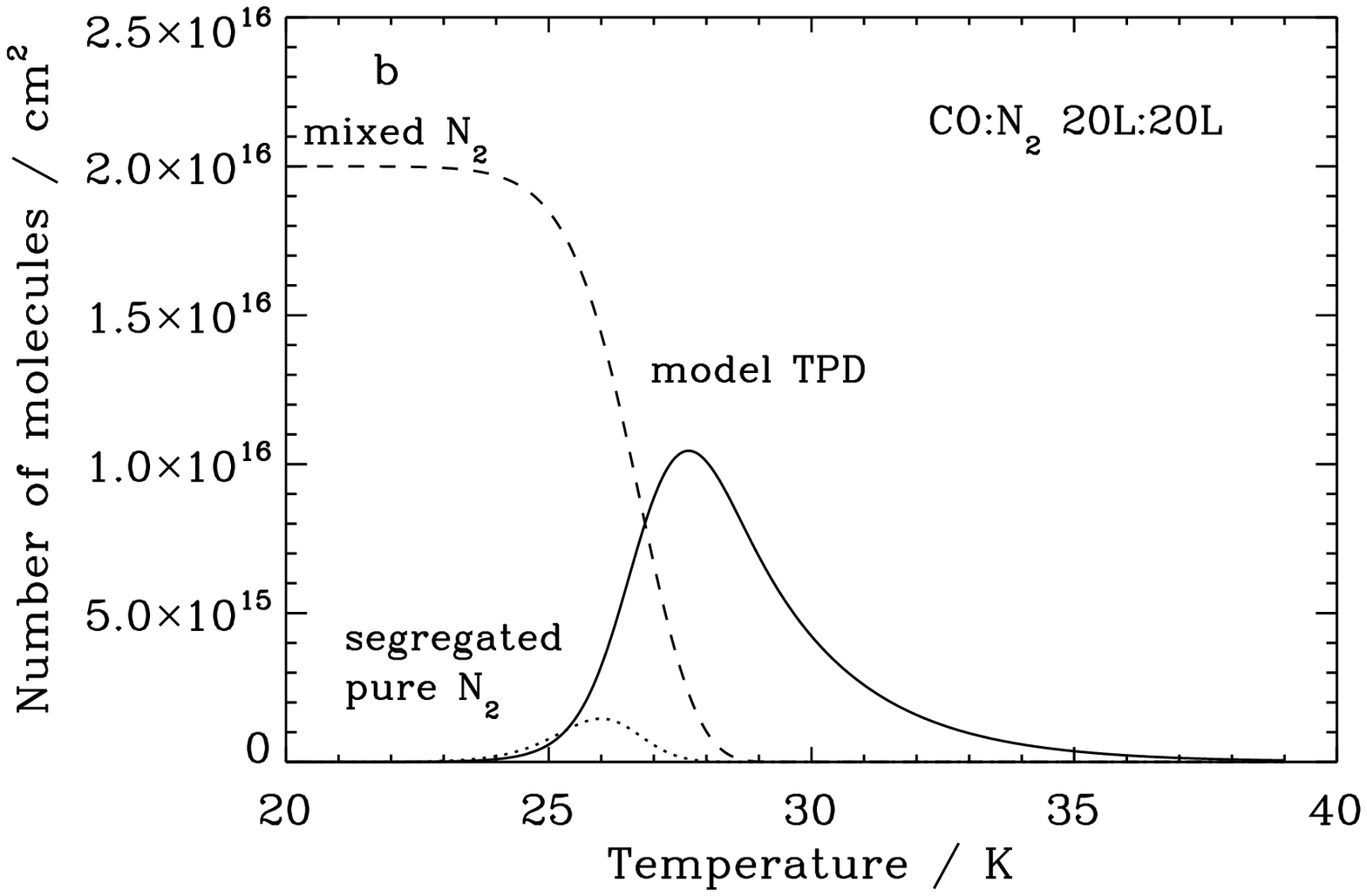}}
   \caption{Model results for the ice and gas phase concentrations as functions of temperature. The number of molecules in pure N$_2$ ice (dotted line), in mixed ice (dashed line), and in the gas phase (solid line) are shown for 20/40 L N$_2$/CO (a) and the 20:20 L CO:N$_2$ (b).}
              \label{ice}
    \end{figure}

A comparison between Fig. \ref{tpd_pure}a and c with Fig. \ref{tpd_pure}b and d clearly shows that the model described here very reasonably reproduces the data of the pure CO and N$_2$ ice system. The leading edges for the CO TPD spectra of the experiment do not quite overlap as perfectly as the model does, probably because CO desorption is close to, but not quite, zeroth order. As was discussed by \citet{collings2003b} the error resulting from this deviation from zeroth order is significantly smaller than all other errors made in astrochemical models. The best-fitted parameters for $E$ are 855 K and 800 K for CO and N$_2$ respectively.

The desorption peaks I and  II observed in the TPD spectra for the layered ices are also well reproduced by the model. The appearance of peak II depends on equations c and d which describe desorption of CO and N$_2$ from the mixed ice phase. From Table \ref{input} it is seen that $\nu$ and $E$ are within the model error-bars identical in each reaction, confirming that N$_2$ and CO co-desorb from this mixed ice phase. Since $E$ from the mixed ice is greater than $E$ from the pure ice, it seems CO and N$_2$ are both more strongly bound in the mixed ice.  The results give a $R_{\rm BE}$ of 0.936 $\pm$ 0.03 for  the pure N$_2$ and CO ices and 1.0 for the mixed ices, within experimental error of paper I. Note that even for layered ices of thicknesses less than 40 L, most N$_2$ desorbs from a mixed ice environment.

Mixing kinetics were confirmed to be zeroth order. The best-fit $E$ value equals 775 K, which is rather close to $E$ found for desorption of pure N$_2$ and indicates that significant mixing only occurs close to desorption of N$_2$, corresponding to the change in RAIR spectra found around 24 K in Sect. \ref{layers}. This behavior is also illustrated in Fig. \ref{ice}a for the 20/40 L N$_2$/CO experiment, where the growth of the mixed ice phase is commensurate with the loss of the pure ice phase and the desorption of the pure N$_2$ layer. For higher ice thicknesses of N$_2$, the competition between mixing and desorption is in favor of desorption from the pure ice layer, leading to the turn-over in peak intensity from peak I and II.

Segregation starts close to the desorption temperature of CO, which is illustrated by Fig. \ref{ice}b for the 20:20 L CO:N$_2$ experiment. This occurs at a higher temperature than the onset of mixing, due to a barrier difference; $E$ equals 930 K for segregation and 775 K for mixing. This difference makes segregation a relatively unimportant process for layered ices. As for mixed ices, however, the segregation rate increases with ice thickness, leading to a larger segregated fraction for higher initial ice thickness, which shifts the TPD peak to lower temperatures. 

\section{Sticking probability}
\label{sticking}

The data presented so far are key to our understanding of CO and
N$_2$ desorption rates in interstellar environments. However,
because the binding energies of CO and
N$_2$ in the solid phase are essentially so similar, this parameter
cannot be the main factor which accounts for the
anti-correlation of N$_2$H$^{+}$ with CO and HCO$^{+}$ in pre-stellar
cores. The freeze-out rate, or a difference in the sticking
probability of each molecule to the grain, may also be
relevant.

Without a molecular beam facility, it is very difficult to quantify
sticking probabilities directly. Nevertheless, during these
experiments, the gas load reaching the mass spectrometer was
monitored during the flow setting for a time period equivalent to
the dosing period (when the substrate was warm) and the entire
dosing period (when the substrate was cold). By combining the
measurements over a range of deposition times and experiments, it is
possible to extract a value for the uptake coefficient. From the uptake coefficient only a lower limit to the sticking probability can be derived since the mass spectrometer signal at low temperatures also includes an unknown fraction of molecules that miss the substrate \citep[for a more detailed explanation of the derivation of the uptake coefficient see][]{fuchs2006}.
The uptake coefficient at surface temperatures of 14 K is given by

\begin{equation}
S(\theta) = \frac{\int N_x^w dt - %a 
\int N_x^c dt}{\int N_x^w dt}
\label{stick_eq}
\end{equation}

\noindent where $\theta$ is the ``thickness'' in L, and $\int N$ is the integrated area under the mass
spectrometer signal for species $x$ during the dosing
period, warm ($w$) or cold ($c$), respectively, which is
directly proportional to the fraction of molecules that do not
stick, i.e. either they never reach the substrate, scatter from the
surface without sticking, or are trapped and desorb on a very short
timescale ($<$ 1 sec). 

However, since the sticking probability is dependent of ice ``thickness'' and ice morphology,
the growth of islands or non-linear thin films during deposition,
such as is observed in these experiments (see Fig. \ref{COgrowth}),
results in the sticking probability changing as a function of ice
``thickness'', tending exponentially (in this case) towards a
constant (lower value) at flat, multilayer ice thicknesses \citep{kolanski2001}.
To determine this
``constant'' $S$-value for CO sticking to CO, N$_2$ sticking to N$_2$,
and N$_2$ sticking to CO, the $S$-values were plotted as a function
of exposure (in L), and fitted to an exponential decay
curve, for every experiment where the final ice morphology was
identical. The asymptotic values of $S$ are given in Table
\ref{stick_tab}. The errors on the uptake coefficients, i.e., the lower limits of the sticking probabilities, arise from
a combination of the reproducibility of the experiments plus the
error bar on the fitted exponential decay curve.

\begin{table}
\caption{Lower limits to the sticking probabilities at 14 K.}
\centering
\begin{tabular}{l|c}
\hline
\hline
System & Sticking probability\\
\hline
CO $\rightarrow$ CO & $\ge$ 0.9 $\pm$ 0.05\\
N$_2$ $\rightarrow$ N$_2$ & $\ge$ 0.85 $\pm$ 0.05\\
N$_2$ $\rightarrow$ CO & $\ge$ 0.87 $\pm$ 0.05\\
\hline
\end{tabular}
\label{stick_tab}
\end{table}

It is clear that at 14 K these values are identical within
experimental error, averaging 0.87 $\pm$ 0.05. The values given in
Table \ref{stick_tab} represent the lower limits to the sticking
probabilities at surface temperatures of 14 K; at higher ice
thicknesses these values will not change, and at lower ice
thicknesses they tend exponentially towards 1. In our experiments,
the non-unity sticking probability may arise because the gases are
dosed effusively into the chamber at 300 K, even though the
substrate itself is at 14 K. 

Based on comparison with other systems it is expected that for a
single molecule incident upon any of these surfaces, the sticking
probability will tend towards 1, particularly as its incident energy
is reduced from 300 to 100 or even 10 K, and the surface temperature
of the ice is reduced to 10 K. The data clearly show that the
relative differences between the $S$-values of CO-CO, N$_2$-N$_2$
and N$_2$-CO are negligible relative to other uncertainties in
astrochemical models, and are certainly not as large as one order of
magnitude, as adopted by \citet{flower2005}. 

\section{Astrophysical implications}
\label{astro}

The model described in Sect. \ref{model} can be refined to simulate the behavior of CO-N$_2$ ices in
astrophysical environments, simply by replacing the heating rate
used in the experiment with an appropriate heating rate for the
astrophysical conditions and removing the pumping reactions.

\begin{figure*}
   \centering
\resizebox{\hsize}{!}{\includegraphics{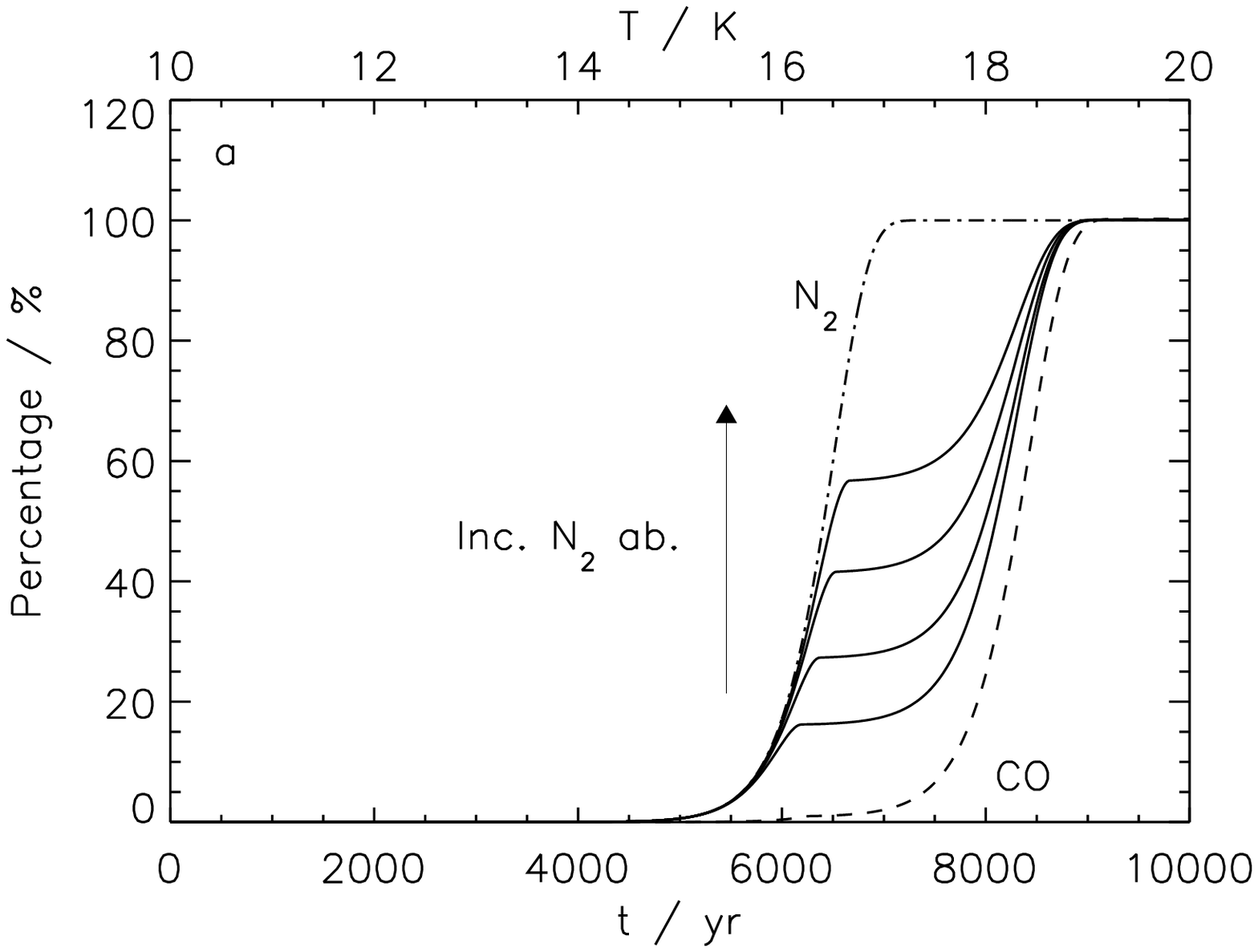}\includegraphics{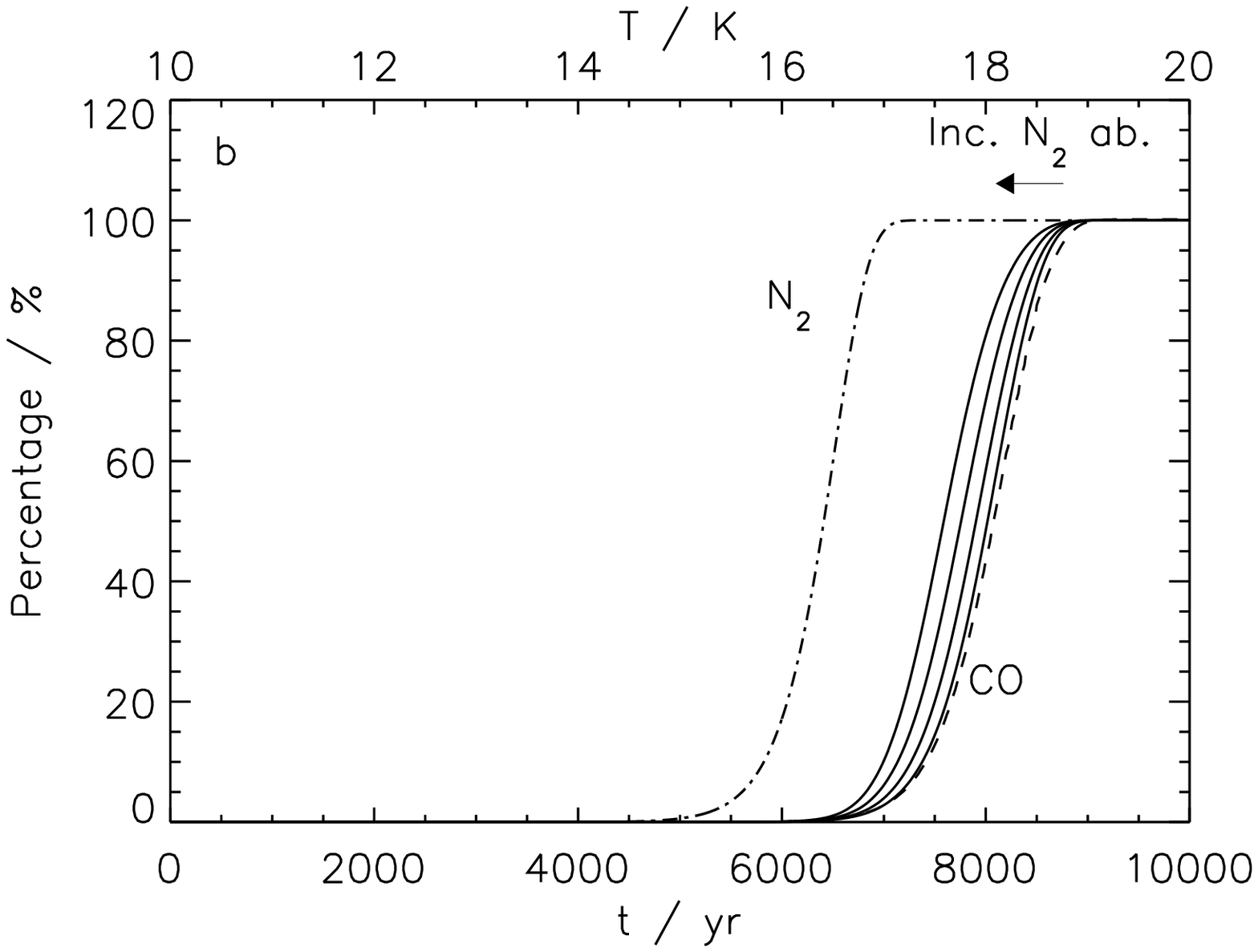}}
\resizebox{\hsize}{!}{\includegraphics{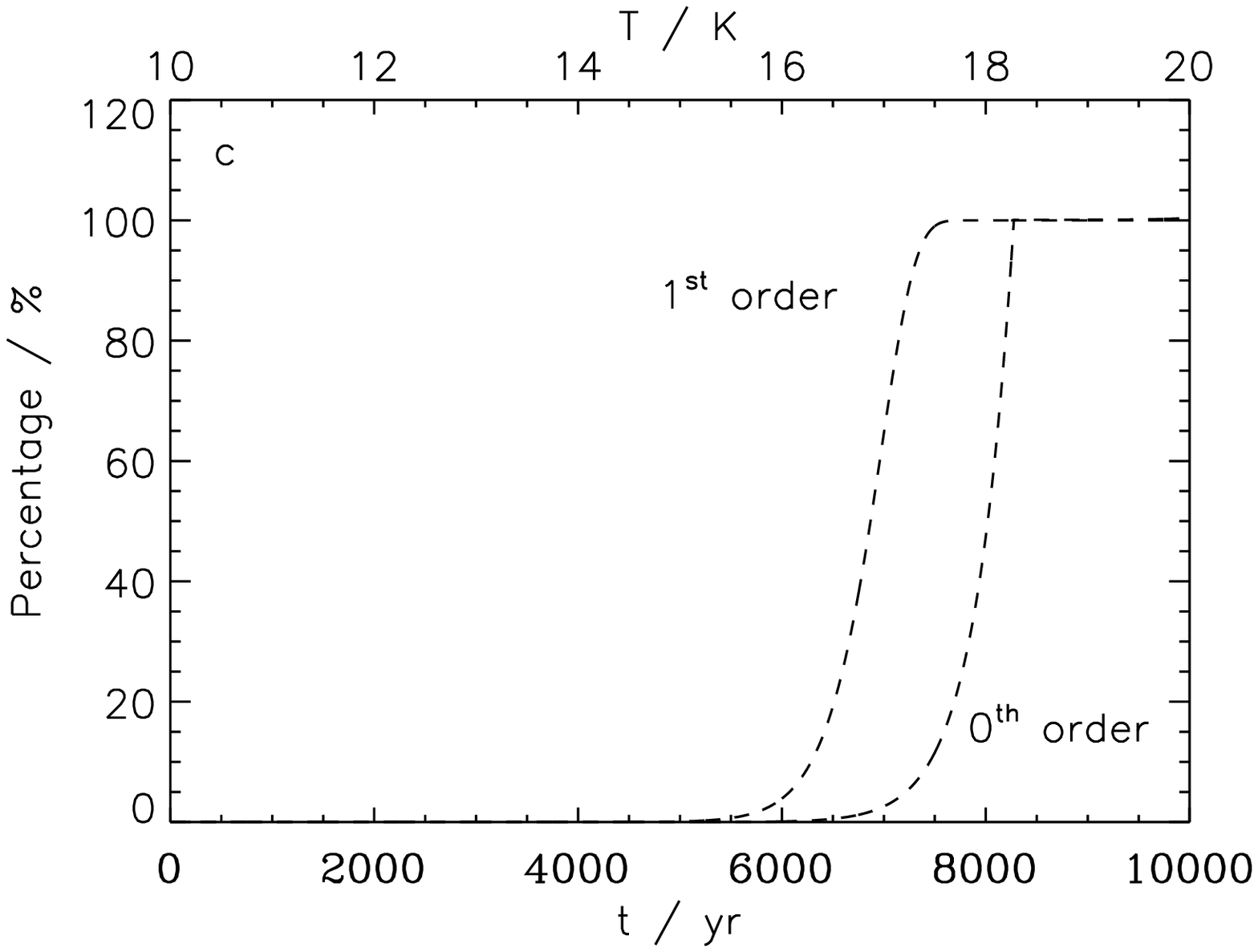}\includegraphics{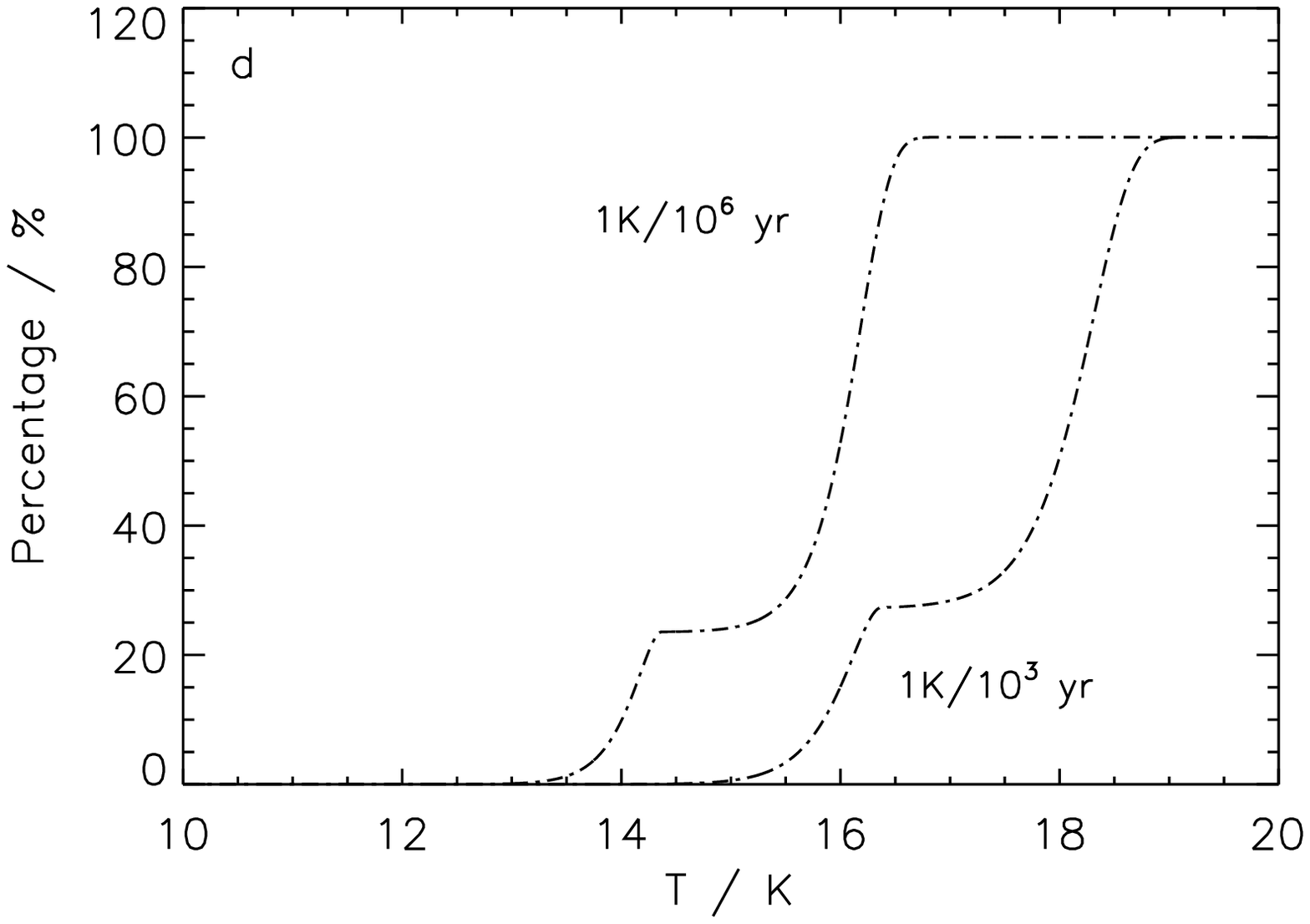}}
   \caption{Astrophysical simulations for heating rates of 1 K/10$^3$
   yr with ice thicknesses ranging from 10 to 80 L for both species. (a) (10-20-40-80 L)N$_2$/(10-20-40-80 L)CO, 1/1 layer, (b) (10-20-40-80 L)N$_2$:(10-20-40-80 L)CO, 1:1 mixed ice; (c) zeroth and first-order desorption for 40 L pure CO, and (d) a
   simulation for N$_2$/CO 20/40 L for heating rates of 1 K/10$^3$ yr and 1 K/10$^6$ yr. N$_2$ desorption from the mixed or layered ices is shown in full, pure N$_2$ in dash-dot, and CO in dashed lines.}
              \label{n2_step}
    \end{figure*}

Fig. \ref{n2_step}a shows output of the astrophysical model
for 1/1 layered N$_2$/CO ices (solid lines), at heating rates of 1
K/1000 yr. This heating rate was chosen because it matches the
timescale over which a newly-formed protostar increases the
temperature in its surrounding envelope from $<$10~K to 20~K
\citep{lee2004}. In addition, the desorption profiles of pure N$_2$
and CO in layered ice are shown on the same plot. Under these conditions, pure
N$_2$ desorbs between 15 and 17 K, $\sim$2 K or 2000 yr earlier than
CO, which desorbs between 17 and 19~K. However, if N$_2$ were to
freeze-out on top of an existing CO-ice layer, the desorption of
N$_2$ takes place in two steps. Only for unrealistically thick ices
of more than 80--120 monolayers does 50\% of N$_2$ desorb as pure
N$_2$. For lower ice thicknesses, N$_2$ desorption from the mixed
environment dominates, and the majority of the frozen-out N$_2$
desorbs with CO. Fig. \ref{n2_step}b shows a very similar plot, but
for 1:1 mixed ices, where the desorption occurs in a single step. As
the total ice thickness increases, i.e. more CO and N$_2$ are
equally frozen out, the desorption profile shifts towards the pure
N$_2$ case, but generally the profile resembles that of pure CO much
more closely than that of pure N$_2$. It is important to note that
the thermodynamics, i.e. $R_{\rm BE}$ of the CO and N$_2$ ice systems have
not been altered in any of these models; the differences
arise entirely from the kinetics of the desorption processes. This
illustrates that it is important to know the initial morphology of
the ice as well the abundance of N$_2$ with respect to CO to make
accurate predictions for the interstellar desorption behavior of N$_2$ compared to CO.

Many astrochemical models use first order desorption kinetics for pure CO
instead of zeroth order kinetics \citep[e.g.,][]{ceccarelli2005}. To get an impression of the magnitude of the
error made by using incorrect desorption kinetics, a simulation
for pure CO desorption from an ice of 40 L was made for both cases
using identical binding energies (see
Fig. \ref{n2_step}c). Clearly, desorption for first order kinetics
occurs $\sim$1 K or 1000 yr earlier, corresponding to an error of 12.5\% on
the desorption timescale. Although this seems a small overall
error, it is 50\% of the time difference between desorption of
pure N$_2$ and CO, so this incorrect treatment could have a comparatively large
effect on the relative desorption behavior of layered ices of N$_2$ and CO. It is also important to notice that CO desorption in pure CO ice is completed $\sim$ 0.5 K earlier than desorption of CO from a mixed or layered ice environment. This is due to the lower surface concentration of CO in a mixed ice environment as was found in the experiments.

In Fig. \ref{n2_step}d, the difference between heating rates of 1 K/10$^3$ yr and 1 K/10$^{6}$ yr is shown for an ice with 20/40 L
N$_2$/CO. The relevance of the faster rate was defined
previously; the slower rate would be appropriate for a cold
pre-stellar core at near constant temperature. It is clear that the
qualitative picture remains the same; N$_2$ desorbs in two steps,
but desorption is complete by 16.5 K for 1 K/10$^{6}$ yr versus at
18.5~K for 1 K/1000 yr, a difference of 2 K for a difference in
heating rate of 10$^3$. One further issue is that at the lower
heating rates a slightly greater fraction of the N$_2$ desorbs from
a mixed ice environment, which implies that the mixing rate becomes
faster relative to the desorption rate. An infinitely slow heating rate of 1 K/10$^{10}$ yr shows the same trend.

The overall conclusion from our experiments is that there are some
subtle differences in the N$_2$ and CO desorption behavior, but that
they are unlikely to fully explain the observed anti-correlations
between N$_2$H$^+$ and CO in pre-stellar dense cores.
Also, any difference in sticking probabilities for CO and N$_2$ is very small, so that other scenarios must be explored to explain
the observations.

So far, H$_2$O ice has been neglected in our studies. The CO-H$_2$O system has been extensively studied by \citet{collings2003b}, who found a binding energy of CO to H$_2$O of 1180 K. \citet{kimmel2001} derive a binding energies of $\ge$ 950 K for N$_2$ on H$_2$O. The combination of these two results in a $R_{\rm BE}$ on H$_2$O of $\ge$ 0.81. Furthermore, \citet{manca2004} and \citet{manca2003} report  a ratio for the condensation enthalpies on H$_2$O of 0.83. Concluding, $R_{\rm BE}$ on H$_2$O for CO and N$_2$ is very close to that found for the binary CO-N$_2$ system. The desorption behavior will thus also be quite similar for CO and N$_2$ in mixed or layered ices with H$_2$O as is observed in the TPD experiments by \citet{collings2004}, where both species desorb in multiple steps. The addition of H$_2$O to the CO-N$_2$ ice system therefore could not significantly alter the conclusions of this paper. A significant difference in binding
energies between CO and N$_2$ could only occur if most of the CO
were residing in a H$_2$O-dominated
environment with N$_2$ in a pure, separate layer on top.  This would
be in contradiction with the observations which show that a large
fraction of the CO is in a pure CO ice layer
\citep{pontoppidan2003}.

Modifications to the gas-phase chemistry are an alternative
possibility to explain the observations.  For example,
\citet{rawlings2002} show that a higher initial H/H$_2$ ratio can
affect the relative N$_2$H$^+$ and HCO$^+$ abundances in cores where
the chemistry has not yet reached equilibrium.  Even in models
including freeze-out, there are regimes of densities and temperatures
where the N$_2$H$^+$ abundance initially rises as CO and N$_2$ freeze out.
Dissociative recombination with electrons then becomes the dominant
N$_2$H$^+$ destruction mechanism \citep{jorgensen2004b}, leading
mostly to NH rather than N$_2$ \citep{geppert2004}. Eventually, this results in N$_2$H$^+$ depletion at high densities and later times. 

\section{Concluding remarks}
\label{conclusion}

New experimental data have been presented for the desorption and
sticking of CO-N$_2$ ice systems.  Furthermore a kinetic model has
been constructed that allows for accurate simulations of the TPD
experiments as well as predictions for the behavior of CO and N$_2$
ices under astrophysical conditions. The key results are:
\begin{itemize}
\item The ratio for the binding energies for N$_2$ and CO in pure ices
is 0.936 $\pm$ 0.03. For mixed ices, the ratio for the binding energies
is 1.0 (see Sect. \ref{results_model}).
\item Desorption of N$_2$ from layered ices occurs in two steps, due
to mixing of N$_2$ with CO. This indicates that for
astrophysically relevant ice abundances, desorption from the mixed
layer dominates, with less than 50\% of N$_2$ able to desorb prior
to CO.
\item In mixed ices, segregation causes the peak temperature for N$_2$
desorption to shift to lower temperatures for higher ice thicknesses, even
though most of the ice desorbs from a mixed ice environment.
Since the onset of segregation is concurrent with desorption, a
single broad desorption step is observed for N$_2$. For
astrophysically relevant ice thicknesses, N$_2$ desorption occurs
close to the CO desorption temperature.
\item The desorption kinetics for CO ice are zeroth order instead of the
commonly adopted first order process, resulting in an error in the desorption timescale of 12.5\%, with a shift to lower temperatures for the first order process. Since this corresponds to 50\% of the difference between N$_2$ and CO desorption, it results in a comparatively large effect on the relative desorption behavior of layered ices of N$_2$ and CO.
\item The lower limits on the sticking probabilities for N$_2$
and CO are found to be the same within experimental error, 0.87
$\pm$ 0.05 at 14~K (see Sect. \ref{sticking}). In reality the
sticking probabilities will be even closer to 1.0 for lower
temperatures.
\end{itemize}

The main conclusion from this work is that the solid-state processes of CO and N$_2$ are very similar under astrophysically relevant conditions.

\begin{acknowledgements}
We are thankful for useful discussions with G. Fuchs, K. Acharyya, and J. J{\o}rgensen. Funding was provided by NOVA, the Netherlands Research School for Astronomy and a NWO Spinoza grant. KO is grateful to the summer undergraduate research fellowship (SURF) program at Caltech for sponsoring her visit to Leiden.
\end{acknowledgements}

%\Online
\begin{appendix}
\section{Comparison between zeroth, first and second order mixing}
\label{Ap_A}

Since there is no direct measurement of the mixing rate  in our experiment, the correct description of mixing kinetics is derived from comparison of models for zeroth, first and second order mixing kinetics with the TPD data. All three mechanisms are physically relevant. Zeroth order can be viewed as a process in which a molecule at the interface between CO and N$_2$ has a certain chance of moving into the overlying layer; the chance for this to occur is completely independent of whether there are 20 or 80 L on top of this molecule and thus the mixing process is ``thickness'' independent. First order mixing would be possible in case mixing of one species with another is independent of the total number of molecules of the other species, i.e. there is no saturation possible. Second-order reactions are possible if the rate of mixing would depend upon both the total number of CO and N$_2$ molecules, since the presence of both molecules is required for mixing. Models for all scenarios were tested in order to determine which most accurately describes the experiments.

\begin{table*}
\caption{Rate equations for the zeroth, first and second order mixing processes.}
\centering
\begin{tabular}{l|ccccc}
\hline
\hline
 & Reaction & Rate equation & $\nu$ & $E$ & $i$\\
& & & (molecules$^{(1-i)}$ cm$^{2(i-1)}$ s$^{-1}$) & (K) & \\
\hline
A & CO(s)+N$_2$(s)$\rightarrow$ CO(mix)+N$_2$(mix) & $\nu_0$e$^{-E/T}$ & 5.0$\times$ 10$^{26 \pm 1}$ & 775 $\pm$ 25 & 0\\
B & CO(s)+N$_2$(s)$\rightarrow$ CO(mix)+N$_2$(mix) & $\nu_1$[CO(s)/N$_2$(s)]e$^{-E/T}$ & 1.0$\times$ 10$^{12 \pm 1}$ & 885 $\pm$ 25 & 1\\
C &  CO(s)+N$_2$(s)$\rightarrow$ CO(mix)+N$_2$(mix) & $\nu_2$[CO(s)][N$_2$(s)]e$^{-E/T}$ & 5.0$\times$ 10$^{-5 \pm 1}$ & 865 $\pm$ 25 & 2\\
\hline
\end{tabular}
\label{order_tab}
\end{table*}

\begin{figure*}
   \centering
\resizebox{\hsize}{!}{\includegraphics{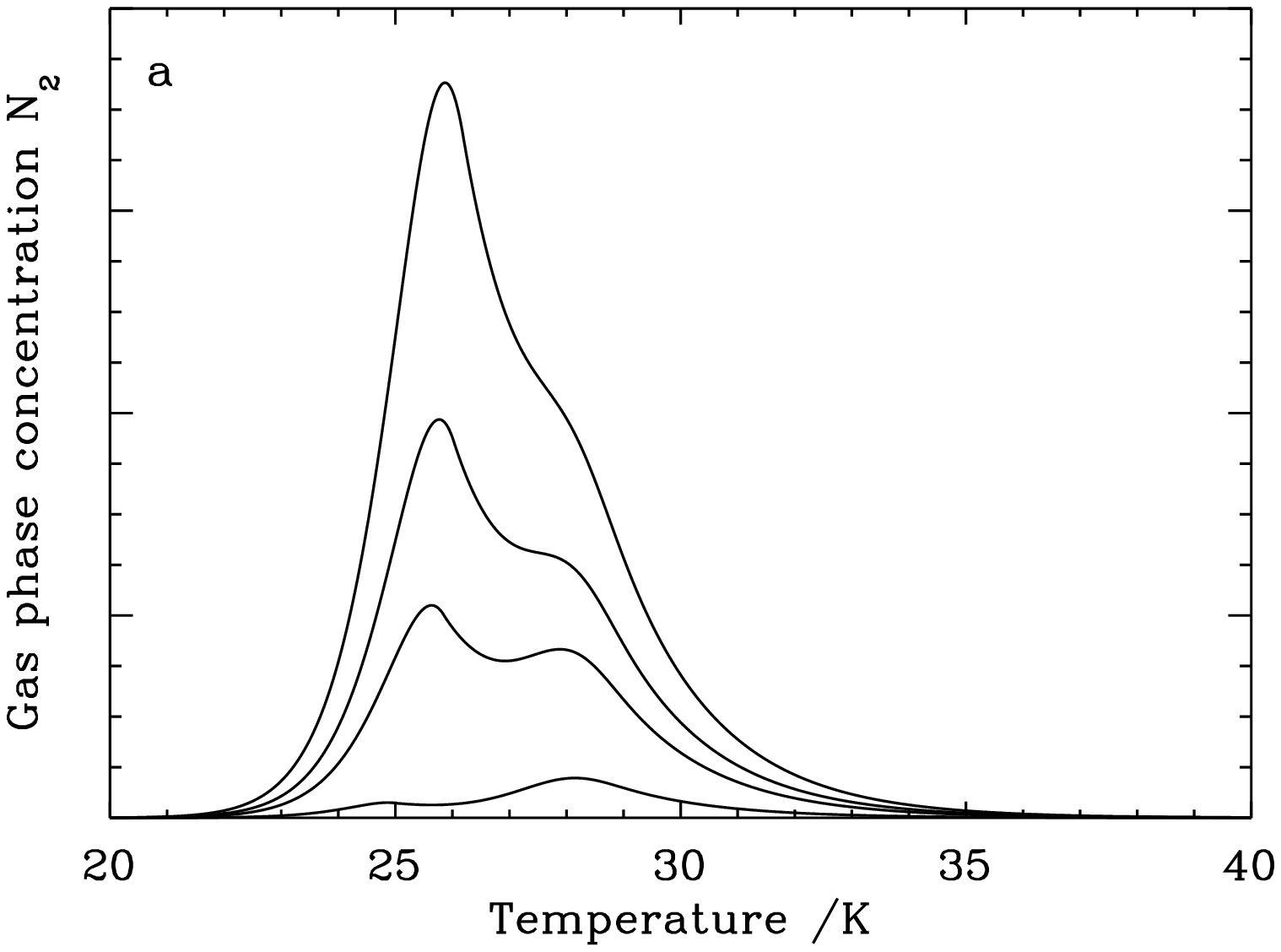}\includegraphics{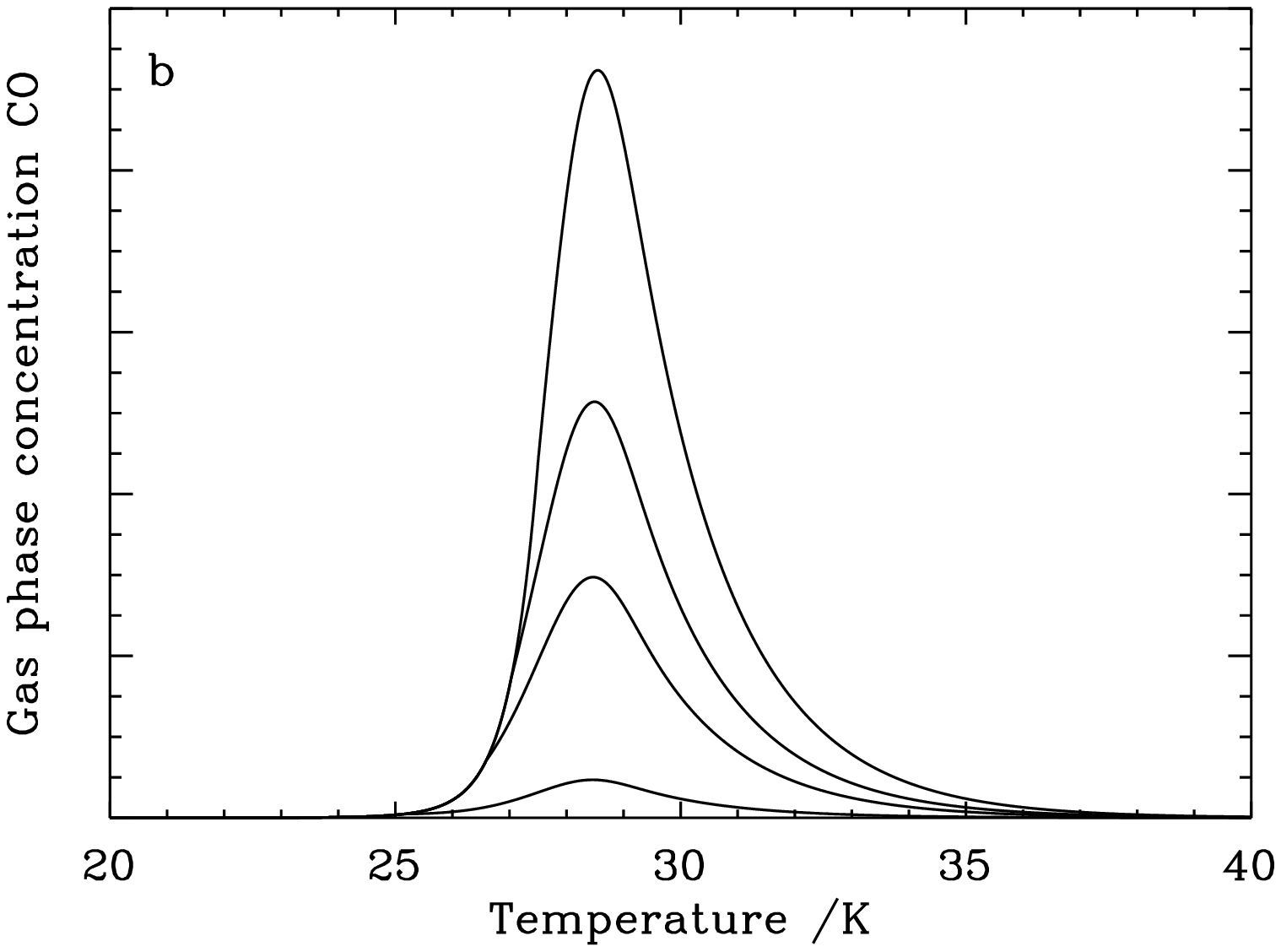} }
\resizebox{\hsize}{!}{\includegraphics{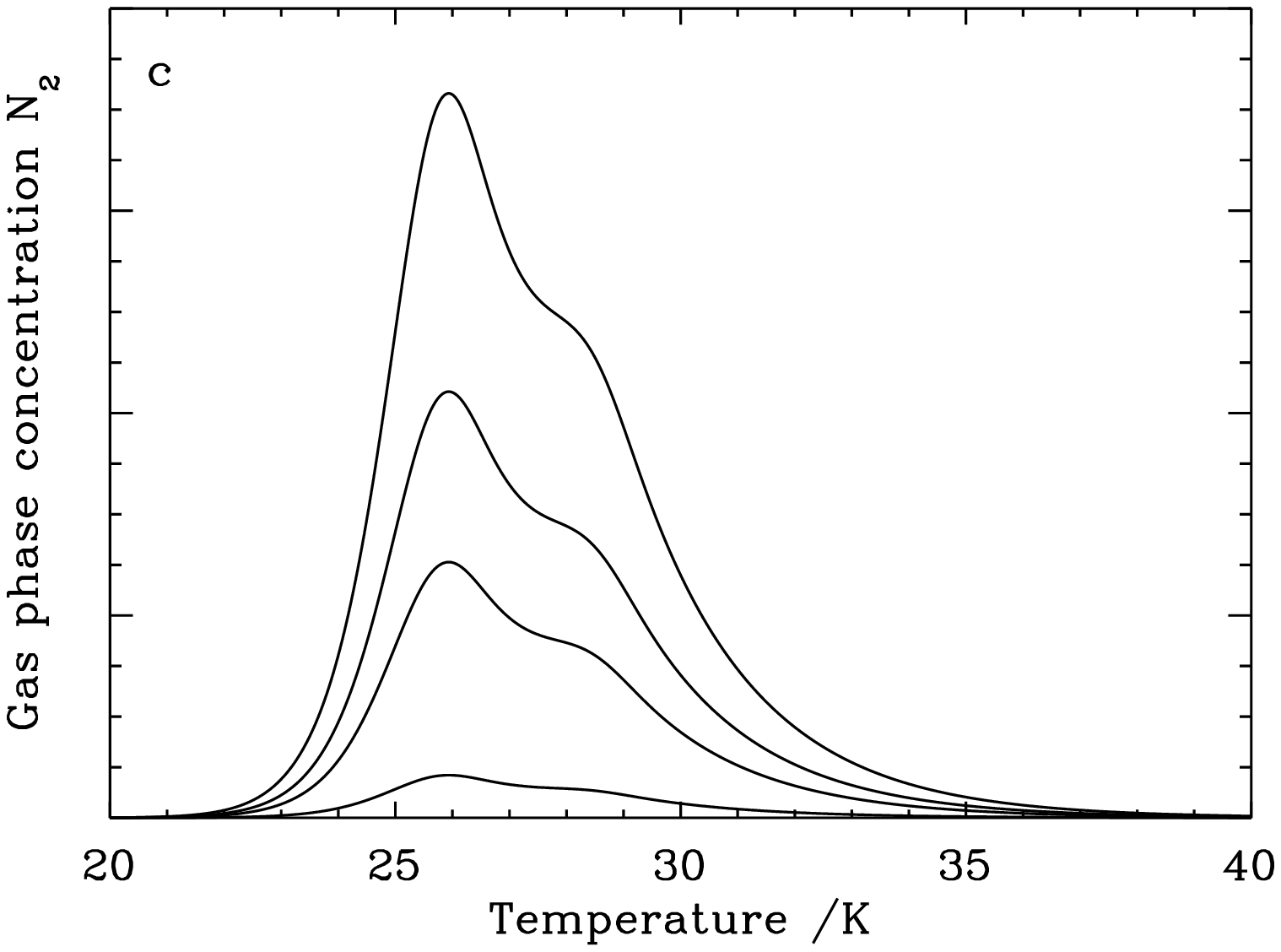}\includegraphics{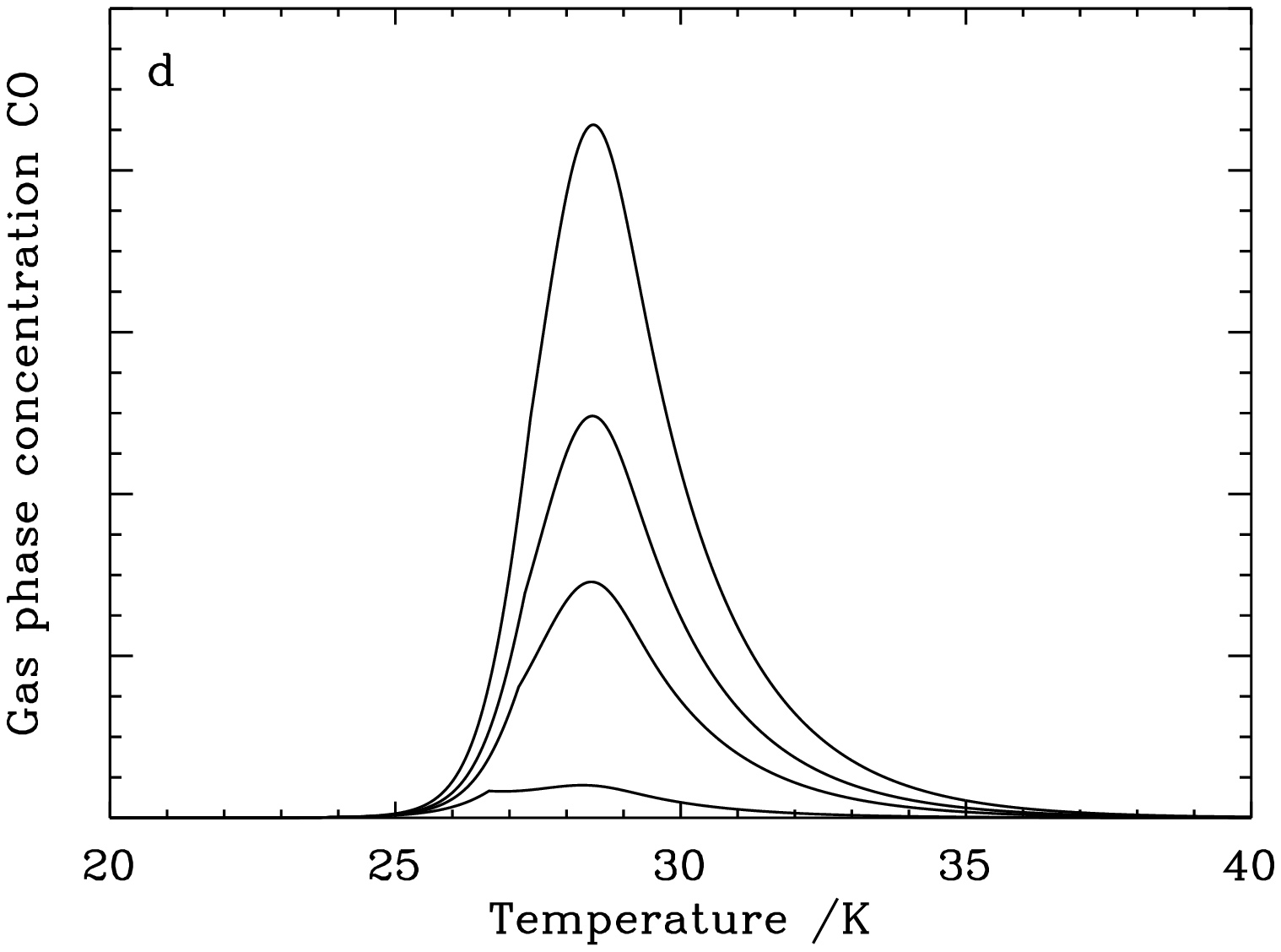} }
\resizebox{\hsize}{!}{\includegraphics{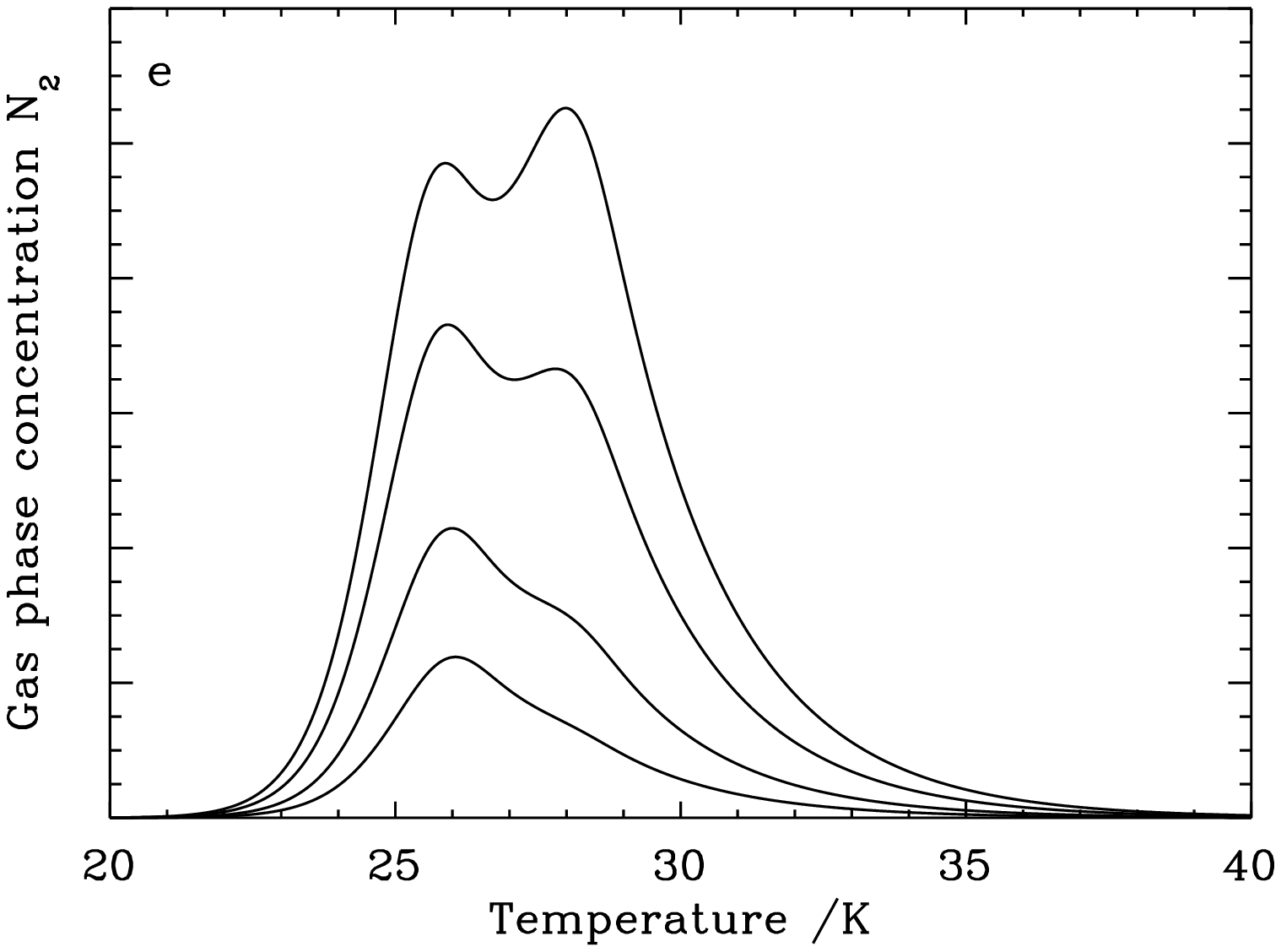}\includegraphics{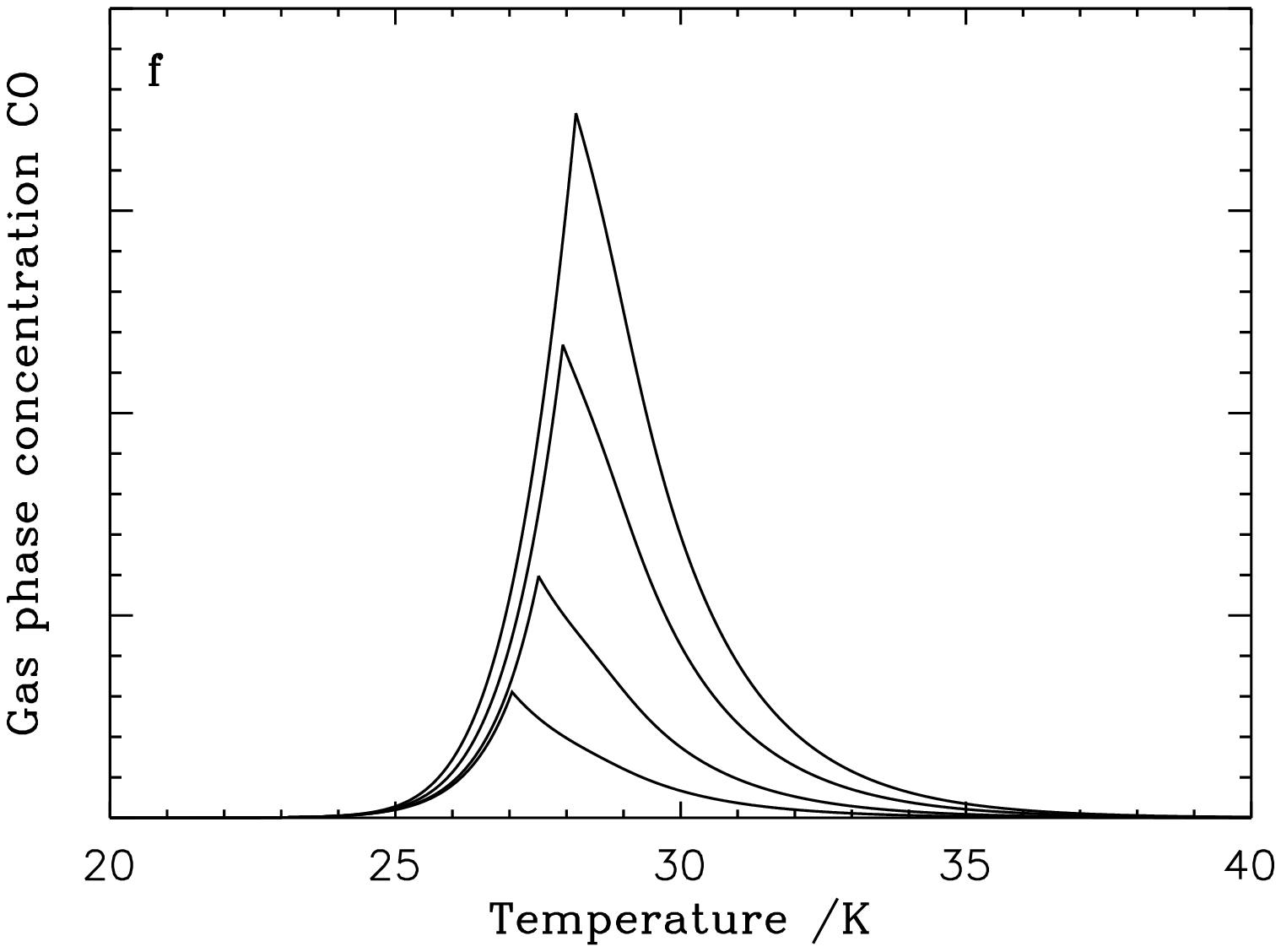} }
   \caption{Comparison between model output for the 1/1 ices using alternative rates for mixing where (a)+(b) have zeroth order mixing kinetics, (c)+(d) first order, and (e)+(f) second order. N$_2$ TPD simulations are shown in (a), (c), and (e); CO TPD simulations are shown in (b), (d), and (f). The N$_2$ model results should be compared with experimental data in Fig. \ref{tpd_overview}a, the CO model results with data in Fig. \ref{CO_TPD}}
              \label{order_mix}
    \end{figure*}

The three different scenarios are shown for 1/1 N$_2$/CO experiments in Fig. \ref{order_mix} with the best fitting parameters in Table \ref{order_tab}. Zeroth order mixing gives rise to a turn-over in the spectrum for the peak intensities with peak II initially being more intense than peak I. This behavior is also observed for the experimental data (see Fig. \ref{tpd_overview}a). The turn-over is due to most N$_2$ molecules mixing unhindered. When the pure ice layer is depleted due to desorption and mixing, mixing stops and the remainder of the molecules desorb from the mixed ice environment. Thus mixing occurs up to higher temperatures with increasing initial ice ``thickness''. Desorption and mixing are therefore competing processes. This behavior is not correctly reproduced by the models for first and second-order mixing (see Fig. \ref{order_mix}c and e). As the initial number of molecules in the layers increases, the number of molecules in the mixed fraction of the ice also increases for first order mixing kinetics (see reaction B in Table \ref{order_tab}). However, this increase is proportional to the number of molecules in the pure layer, resulting in a constant ratio between peak I and II. Second order mixing behaves differently from both zeroth and first-order mixing in that the turn-over is now reversed. This is due to the rate of mixing being proportional to the number of molecules for both species. Thus at low ice ``thicknesses'' the rate is low and both molecules remain mostly pure, whereas for high ice ``thicknesses'' the rate of mixing is very high and all molecules end up in a mixed environment. A comparison between Fig. \ref{order_mix}a, b, and c with Fig. \ref{tpd_overview} shows clearly that the scenario for zeroth order mixing reproduces the experimental data best.

\begin{figure}
\centering
\resizebox{\hsize}{!}{\includegraphics{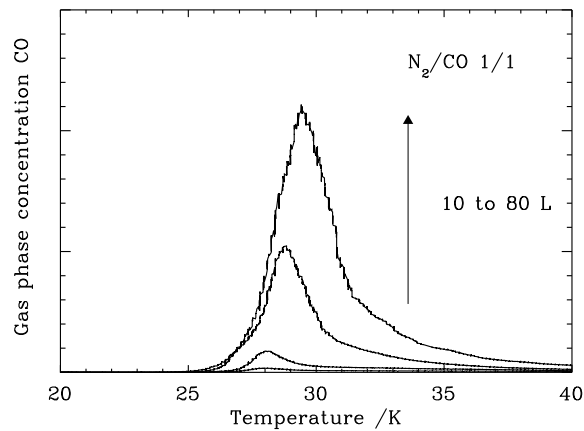}}
\caption{CO TPD spectra for 1/1 (10-20-40-80 L)N$_2$/(10-20-40-80 L)CO.}
\label{CO_TPD}
\end{figure}

The zeroth order mixing mechanism is exemplified by comparison between the CO TPD data output from the model with the experimental data for N$_2$/CO 1/1 (Fig. \ref{CO_TPD}). Second order mixing (Fig. \ref{order_mix}f) produces a CO desorption spectrum that looks zeroth order. For first order mixing, a two-peak structure is observed for lower ``thickness'' ices and desorption is dominated by first order kinetics for higher ice ``thicknesses'', which is not observed in the experimental data. Zeroth order mixing kinetics are, however, able to predict the increasing overlap for the leading edges plus the slight broadening of the TPD profile with respect to the pure CO TPD spectra in Fig. \ref{tpd_pure}a with increasing ice thickness observed in the experimental data fairly well.

\end{appendix}

\end{document}